
\def\varep{\varepsilon}
\def\ltgt{<>}

\def\ga{ \gamma }
\def\mpe{ m_p/m_e }
\def\flu{\ergs\cmsqi}

\def\ka{ \kappa}
\def\la{ \lambda}

\def\Kpc{{\rm\,kpc}}

\def\cm{{\rm\,cm}}

\def\ga{ \gamma}
\def\etal{{\it et~al.}}

\def\halfspace{\baselineskip=12pt plus .1pt}

\def\papersize{\magnification=1200}  

\font\tfont=cmmib10
\newfam\vecfam

\textfont\vecfam=\tfont \scriptfont\vecfam=\seveni
\scriptscriptfont\vecfam=\fivei

\parindent=40pt
\settabs 7 \columns
\tolerance=1600
\parskip 1ex
\def\foolit{\ifnum\pageno > 1 \number\pageno\fi}
\raggedbottom
\def\frac#1#2{{#1\over#2}}
\def\Mesz{M\'esz\'aros\ }
\def\Pacz{Paczy\'nski\ }

\def\ctl{\centerline}

\def\gbr{\goodbreak\noindent}

\def\ref{\par \noindent \hangindent=2pc \hangafter=1 }
\def\etal{{\it et~al.\ }}
\def\mathnew{\mathsurround=0pt}
\def\simov#1#2{\lower .5pt\vbox{\baselineskip0pt \lineskip-.5pt
	\ialign{$\mathnew#1\hfil##\hfil$\crcr#2\crcr\sim\crcr}}}
\def\simg{\mathrel{\mathpalette\simov >}}
\def\siml{\mathrel{\mathpalette\simov <}}
\def\lambdabar{\mathrel{\lower 1pt\hbox{$\mathchar'26$}\mkern-9mu
        \hbox{$\lambda$}}}

\def\ssk{\vskip 1ex\noindent}
\def\msk{\vskip 2ex\noindent}
\def\bsk{\vskip 3ex\noindent}

%
\def\cm{~\rm{cm}}

\def\s{~\rm{s}}
\def\si{~ {\rm s}^{-1} }

\def\cmsqi{~ {\rm cm}^{-2} }
\def\cmcui{~ {\rm cm}^{-3} }

\def\keV{~\rm{keV}}
\def\KeV{~\rm{keV}}
\def\MeV{~\rm{MeV}}
\def\GeV{~\rm{GeV}}
\def\TeV{~\rm{TeV}}

\def\erg{~\rm{ergs}}
\def\ergs{~\rm{ergs}}

\def\G{~\rm{G}}

\def\Hz{~\rm{Hz}}

\papersize
\halfspace
%
\ctl{\bf SPECTRAL PROPERTIES OF BLAST WAVE MODELS OF}
\ssk
\ctl{\bf GAMMA-RAY BURST SOURCES}
\bsk\bsk
\ctl{ P. \Mesz$^1$, M.J. Rees$^2$ and H. Papathanassiou$^1$}
\bsk
\ctl{$^1$ Pennsylvania State University, 525 Davey Lab., University Park, PA
16802}
\ctl{$^2$ Institute of Astronomy, Madingley Road, Cambridge CB3 0HA, England}
\bsk
\ctl{Ap.J., Received:~11/19/93;~Accepted:~~~~~~~~}
%
\def\E{ E_{51} }
\def\Ec{ E_{51} }
\def\Eh{ E_{41} }
\def\Ed{ E_{39} }
\def\th{ \theta}
\def\ka{ \kappa}
\def\kas{ \kappa_{sy}}
\def\kai{ \kappa_{ic}}
\def\la{ \lambda}
\def\ga{ \gamma }
\def\Ga{ \Gamma }
\def\Gas{ \Gamma_{sy} }
\def\Gai{ \Gamma_{ic} }
\def\eth{ \eta_3 }
\def\etw{ \eta_2 }
\def\n{ n_0 }
\def\nh{ n_{-3} }
\def\ng{ n_{-3} }

\def\froz{  frozen-in~ }

\def\equip{ turbulent }
\def\sync{ synchrotron~}
\def\syt{ {syt}}
\def\ict{ {ict} }
\def\usy{ u_{sy} }
\def\ub{ u_B }
\def\tsy{ t_{sy}}
\def\tic{ t_{ic}}

\def\tex{ t_{ex}}
\def\mc2{ m_e c^2}

\def\mpe{ m_p/m_e }

\def\flu{\ergs\cmsqi}
\def\ende{\ergs\cmcui}
\bsk
\ctl{\bf Abstract}
\bsk
We calculate the spectrum of blast wave models of gamma-ray burst sources,
for various assumptions about the magnetic field density and the relativistic
particle acceleration efficiency. For a range of physically plausible models
we find that the radiation efficiency is high, and leads to nonthermal
spectra with breaks at various energies comparable to those observed in the
gamma-ray range. Radiation is also predicted at other wavebands, in particular
at
X-ray, optical/UV and GeV/TeV energies. We discuss the spectra as a function of
duration for three basic types of models, and for cosmological, halo and
galactic
disk distances. We also evaluate the gamma-ray fluences and the spectral
characteristics for a range of external densities. Impulsive burst models at
cosmological distances can satisfy the conventional X-ray paucity constraint
$S_x/S_\gamma \siml$ few percent over a wide range of durations, but galactic
models
can do so only for bursts shorter than a few seconds, unless additional
assumptions
are made. The emissivity is generally larger for bursts in a denser external
environment, with the efficiency increasing up to the point where all the
energy input is radiated away.
\bsk
%
\ctl{\bf 1.~ Introduction}
\bsk
In most gamma-ray burst (GRB) scenarios, the very high initial radiation
density leads  to a relativistically expanding fireball. This is unavoidable if
the GRB occur at cosmological distances (\Pacz, 1986, Goodman, 1986), and is
also expected in many  galactic models. For pair-dominated fireballs, an
energetic but very short burst is obtained, whereas in  the more common
baryon-dominated fireballs most of the initial  thermal energy gets converted
into kinetic energy  before it can be radiated away (Cavallo and Rees, 1978,
\Pacz, 1990,  Shemi and Piran, 1991), and only a weak burst is expected.
However the kinetic  energy of the coasting baryons can be reconverted into
radiation when the ejecta are decelerated by the external medium (Rees and
\Mesz, 1992, \Mesz and Rees, 1993). This  solves the total energy problem;
it also results in an observable burst with a longer timescale since this is
then determined by the dynamic timescale at
the deceleration radius. For magnetically dominated fireballs the relativistic
outflow is predominantly in the form of Poynting flux (e.g. Narayan,  \Pacz and
Piran, 1992, Usov, 1992). In all cases, a blast wave moves ahead of the contact
discontinuity, while a reverse compression shock moves into the ejecta,
randomizing the  directed kinetic energy of the fireball. These shocks can
accelerate particles which, in  the presence of magnetic fields (either
frozen-in or created by turbulent instabilities)  lead to a radiation burst
whose total energy is comparable to the initial energy input and whose
time-delayed duration in the observer frame is comparable to that observed.

The radiation in these models depends on the efficiency of particle
acceleration behind  the shocks and on the magnetic field strength. The details
of the particle acceleration process in relativistic shocks are  uncertain, as
are the mechanisms determining the magnetic energy density. We therefore
explore various possibilities. Our first aim is to discover what range of
parameters (e.g. expansion bulk Lorentz factors,
external density) result in gamma-ray
production  with an efficiency and timescale appropriate for the observed
bursts. Secondly, we consider the spectrum of the resulting radiation, and
particularly the extension of this spectrum towards both softer and harder
energies. This is interesting, among other reasons, because of the well-known
`X-ray paucity constraint' on some observed bursts. In our models, the spectra
generally consist of two or more components, and are therefore not a simple
power-law. The relative intensities in the different bands depend on the
frequencies at which the breaks between these components occur. Observations
(or even upper limits) in the X-ray band, or in the optical and UV, can
therefore help to pin down the parameters of the model.

Our relativistic blast wave models were originally applied to cosmological
bursts. However, the same phenomenon would occur, on a smaller scale, when
relativistic plasma released, e.g., by magnetospheric disturbances on a
galactic neutron star runs into the interstellar medium (Begelman, \Mesz and
Rees, 1993). Our considerations are therefore also relevant for galactic models
of gamma-ray burst sources, if one assumes that sufficient sources with the
appropriate energy are available. We stress, however, that the blast wave burst
model is insensitive to the specific energy input source. It is almost
irrelevant
whether the initial energy source is a merging neutron  star binary, a failed
supernova, a collapse-induced giant magnetic flare, or any other  process with
the right total energy. In this paper we examine the efficiency, duration  and
spectral properties of both galactic and extragalactic GRB blast wave models.
\bsk
\ctl{\bf 2.~ The Shock Burst Scenario}
\bsk
2.1~ Standard Parameters and Basic Picture
\msk
For a fireball moving into an external medium characteristic of a  cosmological
scenario, the typical parameters are $n_{ext}\sim 1~n_0\cmcui,~E_o \sim
10^{51}\Ec\ergs,~\eta\sim 10^3\eta_3$, where $E_o$ is the initial fireball
energy, $\eta=E_o/M_o c^2$, $M_o$ is the entrained baryon mass and $n_{ext}$ is
the density of the external medium into which the fireball expands. The input
total energy $E_o$  may be initially dominated by either a radiation/pair
mixture or by magnetic fields, depending on the initial conditions of the
burst. The corresponding numbers for a galactic  scenario would be $n_{ext}\sim
10^{-3}\ng \cmcui$, $E_o\sim 10^{41}\Eh\ergs$, $\eta\sim  10^2\eta_2$ (halo)
and $n_{ext}\sim 1~n_0\cmcui,~E_o=10^{39}\Ed\ergs,~\eta= 10^2\etw$  (disk).
The bulk Lorentz factor initially grows as $\Gamma(r)\propto r$, and eventually
saturates to the value $\Gamma_f\sim \eta$, when the internal energy per baryon
becomes  non-relativistic. Acceptable models for impulsive fireballs generally
require $\eta\siml  10^3-10^4$ (e.g. \Mesz and Rees, 1993) in order for the
burst energy and timescale not to  be too low.  The observer-frame deceleration
radius $r_d$, and the  corresponding comoving-frame
expansion timescale $t_{ex}$ and density of the ejecta material, are, for these
three scenarios,
$$
r_d\sim           10^{16}\n^{-1/3}\E^{1/3}\th^{-2/3} \eth^{-2/3}\cm \sim
        2.2\times 10^{14}\nh^{-1/3}\Eh^{1/3}\th^{-2/3}\etw^{-2/3}\cm \sim
$$ $$
        4.6\times 10^{12}\n^{-1/3}\Ed^{1/3}\th^{-2/3}\etw^{-2/3}\cm
\eqno(2.1.1)
$$ $$
t_{ex}\sim           10^3 \n^{-1/3}\E^{1/3}\theta^{-2/3}\eta_3^{-5/3}\s \sim
           2.2\times 10^2 \nh^{-1/3}\Eh^{1/3}\th^{-2/3}\eta_2^{-5/3} \s \sim
$$ $$
                      4.6 \n^{-1/3}\Ed^{1/3}\th^{-2/3}\etw^{-5/3} \s
{}~,\eqno(2.1.2)
$$ $$
n_d \sim 10^6 \n \eta_3^2 \cmcui \sim 10^2 \n\etw^2\cmcui \sim
                                   10^{-1}\nh\eta_2^2 \cmcui~,\eqno(2.1.3)
$$
where $\eta$ is the bulk Lorentz factor at the start of deceleration and
$\theta$ is  the opening half-angle (if not spherical) along which the fireball
expands.  The observed (lab frame) burst duration is $\eta$ times shorter than
the comoving  dynamic time $t_{ex}$, i.e.,
$$
t_L \sim      1~ \n^{-1/3} \E^{1/3} \theta^{-2/3} \eta_3^{-8/3} \s
       \sim 2.2~\nh^{-1/3}\Eh^{1/3} \th^{-2/3}\etw^{-8/3} \s
$$ $$
        \sim 4.6\times 10^{-2} \n^{-1/3}\Ed^{1/3}\th^{-2/3} \etw^{-8/3} \s ~.
\eqno(2.1.4)
$$
Generally the deceleration of the ejecta occurs after the bulk Lorentz factor
has  saturated, $r_d \gg r_s\sim \eta r_o$, and also after $r_d \gg r_b\sim
\eta^2  r_o$ (\Mesz, Laguna and Rees, 1993), so the lab-frame  expansion time
$t_L\sim r_d c^{-1}\eta^{-2} \gg r_o/c\sim  10^{-4}r_6\s$ is of the order of
seconds, and occurs well after the  pairs of the
ejecta drop out of equilibrium. (The ejecta become optically thin to baryonic
electrons  at a radius smaller than the  deceleration radius,
when a weak and very brief burst of $t_t\sim r_o/c\sim  10^{-4}\s$ is produced
which precedes the shock burst by a time of order  $\sim t_L$).

A strong relativistic blast wave moves ahead of the fireball. This blast wave
decelerates as it sweeps up external matter; a reverse shock wave starts to
move into the ejecta, and becomes marginally relativistic when the expanding
ejecta reach a radius  $\sim r_d$. New pairs may be formed in these shocks,
whose maximum effect on the opacity can be evaluated by estimating the comoving
compactness parameter at the deceleration radius, assuming all the kinetic
energy available is converted into radiation above the pair formation
threshold. This is  $\tau_\pm \sim L\sigma_T/ m_e c^3 \Delta R \sim n_\pm
\sigma_T \Delta R \sim  10^{-3}\E^{1/3}\n^{2/3}\theta^{4/3}\eta_3^{1/3} \sim
10^{-8}\Eh^{1/3}\ng^{2/3}\theta^{4/3}\eta_2^{1/3} \ll 1$, i.e. both shocks are
optically thin to pair formation (and also to baryonic electrons).  The
spectrum of the radiation produced by the shocks is therefore virtually assured
to be  {\it non-thermal}, since the shocked electrons are relativistic and emit
in an optically thin environment. Indeed, anything other than a non-thermal
spectrum  would be surprising under these circumstances.

The foregoing discussion, which treats the primary energy production as
instantaneous, is self-consistent provided that the fireball is created in a
time
short compared with $t_L$. The actual energy release will, of course, really
have a
finite duration, and the fireball's early development will depend on the
time-history of the energy production and on the behavior of $\eta$. An
interesting possibility, especially relevant to `cosmological' models
which invoke a coalescing compact binary,  is that
the primary energy emerges as a magnetically-dominated wind lasting for up to a
few seconds (e.g. Narayan \etal 1992; Usov 1992,1993) . Some gamma rays may
come
directly from the inner part of the wind (just as  they can come  from the
'thinning'
stage of an impulsive fireball). However, just as in the impulsive
case, the wind will inflate a cavity, piling up external matter behind a blast
wave, and may thereby convert its energy into gamma rays more efficiently.

To  explore fully the consequences of non-impulsive (and possibly
time-dependent) energy release, one would need to introduce further parameters.
It is premature to do this in an elaborate way. It is, however, worthwhile and
interesting to consider a simple (and physically plausible) illustrative case
when the wind  itself is unloaded and almost purely electromagnetic, so that
its speed is not significantly below $c$. Suppose this wind turns on at $t=0$,
and maintains a steady isotropic  energy flux $L$  for a time $t_w$. It will
inflate a spherical cavity of radius $r$. If the  contact discontinuity at the
cavity boundary expands with a Lorentz factor $\Gamma(r)$, the wind pressure,
in the frame of the contact discontinuity, is proportional to $L/r^2 \Gamma^2$.
This would be balanced by the pressure of the shocked material between the
blast wave and the shock discontinuity, which scales as $n_{ext}\Gamma^2$. The
expansion Lorentz factor will be that for which these two forces balance, and
this implies $\Gamma\propto r^{-1/2}  n_{ext}^{-1/4}$.

When the wind ceases, at a time $t_w$ after it switches on (measured in the
observer frame), the part of the blast wave directed towards the observer will
have reached a radius $r=\Gamma^2 c t_w$ and $\eta$ will have fallen (owing to
the  `loading' by swept-up matter) to $\sim (L/4\pi c^2 t_w^2 n_{ext})^{1/8}$.
At that stage,  the configuration would be essentially equivalent to a fireball
where $E = Lt_w$ and  where $\eta$ has the value defined above. For these
values, we find (as consistency  requires) that the value of $t_L$ given by
(2.1.4) is indeed equal to $t_w$.

Although we present our subsequent discussion in terms of impulsively-produced
pair, baryon or magnetically dominated fireballs, from the above discussion it
applies equally to the corresponding wind models at the phase $t \simeq t_L$
when
they are releasing most of their energy. The radiation from an
impulsively-produced
homogeneous fireball rises to a peak after a time
$\sim r_d c^{-1}\eta^{-2}$ and then fades. The details of the initial rise
would be
slightly different in the wind case as indeed they would for a (more realistic)
inhomogeneous fireball that could not be characterised by a single value of
$\eta$. In this paper we are concerned with the intensity and spectrum of the
radiation emitted near the peak; the key parameters are then
$\eta, \ (\left(E/\theta^2\right)$ and $n_{ext}$.
\bsk
2.2~ Shock Structure, Radiation Densities and Timescales
\bsk
A crucial question is whether the radiative efficiency of the shock(s) is
sufficiently  high, and this depends on the magnetic field. The synchrotron
radiative efficiency of  the deceleration blast wave and of the reverse shock
are near unity if turbulent  instabilities at the shocks lead to magnetic
fields not too far below equipartition. Shocks are unstable (e.g. Ryu and
Vishniac, 1992), which should result in turbulence, and turbulent field growth
seems to be efficient in observed supernova remnant and radio  source jet
shocks. Alternatively, the ejecta may be magnetically dominated,
or at least contain magnetic fields which {\it initially} contribute
a fraction $\xi$ of the dynamical equipartition value,
and  subsequently remain frozen-in, so $B^2\propto V^{-4/3}$, where $V$ is the
comoving volume. In this case the deceleration-induced reverse shock is
expected to  have high inverse Compton (IC) radiative efficiency (\Mesz, Laguna
and Rees, 1993), for  moderate values of the electron Lorentz factor.
Previously, we had assumed that the  IC scattering occurs on the same electrons
that are responsible for the synchrotron photons, an assumption which is
relaxed here. In both the turbulent generation and the frozen-in scenario, the
main uncertainties are the value of the magnetic field at the deceleration
radius, and the value of the typical Lorentz factor of the radiating electrons,
the latter being dependent on the type of acceleration mechanism at the shock.
However,  under the minimal assumption that electrons are accelerated to a
power law distribution,  it is possible to derive reasonable lower limits to
the synchrotron and IC efficiencies  under various parametrizations of the
field strength and the minimum electron Lorentz factor.

The detailed structure of the shocks will depend on the magnitude of the field
and on the efficiency for accelerating electrons to relativistic energies. The
strength and  Mach number of the blast wave and the reverse shock should be
different, as well as  the seed field density, so in principle the field growth
and acceleration efficiency  could be different in both. The importance of the
synchrotron and IC mechanisms will  depend on the strength of the magnetic
field behind the blast wave and the reverse shock,  and several possible
emission region configurations can be considered.

One extreme case is that where the ejecta have a frozen-in magnetic field left
over from the explosion phase, at which time its energy density made up a
fraction $\xi$ of the total initial radiation density. In this case the
magnetic field strength in the reverse-shocked region at the  deceleration
radius $r_d$ is $ B_{df}\sim 0.4\xi^{1/2}\E^{-1/6} \n^{2/3} \eta_3^2~\G~$. On
the other hand, if the shocks are subject to instabilities, leading to
magnetic field growth which reaches some fraction $\lambda$ of the
equipartition value with the corresponding post-shock energetic particles (as
in supernova remnants, or radio sources), then the field at the deceleration
radius is $ B_{de}\sim 4\times 10^2 \n^{1/2}\eth \la^{1/2} \G$. Notice that the
equipartition (or sub-equipartition) field value is the  same (if $\la$ is the
same) behind both the blast wave and the reverse shock, because the particles
behind both must be in pressure equilibrium with each other. The
corresponding magnetic energy densities ($\erg\cmcui$) are
$$
u_B\sim\cases{ 6.6\times 10^3\n\la\eth^2 \ergs\cmcui~,& (\equip );\cr
          7\times 10^{-3}\n^{4/3}\E^{-1/3}\xi\eth^4 \ergs\cmcui~,& (\froz).\cr}
\eqno(2.2.1)
$$
The parameter $\la$ could be different in the blast wave and reverse shock (in
the
equipartition case), due to the different physical conditions, while the
frozen-in case refers only to the reverse shock.

The acceleration mechanism is probably more efficient when the shock is strong
and relativistic. Because the external medium is relatively cold and devoid of
strong fields its sound speed is low, and the blast wave is certain to be
strong and ultrarelativistic since the bulk Lorentz factor $\Gamma$ is large.
If the
acceleration mechanism does not depend on the presence of fields stronger than
the
minimal seed field expected in the interstellar medium, an electron power law
can be
expected to form behind the blast wave. However the strength of any synchrotron
radiation from these electrons does depend on the field strength. Thus
synchrotron
emission from the blast wave will be generally negligible for the fields
$B\sim 10^{-6}\G$ typical of normal ISM, but will be important if turbulent
field growth occurs. IC scattering, however, only requires the  presence of
relativistic electrons, and may be expected to be important under a wide range
of circumstances in the blast wave.

The reverse shock which moves into the ejecta encounters gas which is generally
hotter  than the external ISM, even after adiabatic cooling.  For a
matter-dominated fireball, the sound speed is higher and the reverse shock
becomes at most marginally relativistic. Similarly, for a  magnetically
dominated fireball the Alfv\'en speed is high. However, in either case, the
fireball has a large outward Lorentz factor with respect to the contact
discontinuity, and the reverse shock may be strong, even if not as highly
relativistic as the blast wave. Particles are plentiful in the matter-loaded
fireball, and in the magnetically dominated case particles would be expected
due to entrainment or due to sweeping up of neutrals from the external medium,
which could be accelerated by the reverse shock. However, because both the
field strength and the particle supply are different as well as the shock
strength, it would not be surprising if the blast wave and reverse shock
differed in their ability to accelerate particles and radiate, unless the
interface were unstable and mixing occurred on short timescales.

To cover the possibilities discussed above, we consider several basic shock
models. The simplest is the ``frozen-in" (F) model, which assumes the existence
of
fields in the ejecta at some fraction $\xi$ of the (initial) equipartition
value, and no turbulent field growth in either the blast or reverse shock.
Another model is the ``turbulent" (T) model, where turbulent field growth is
assumed to occur in both shocks leading to fields with an energy density which
is a fraction $\la$ of the particle energy in the shock ($\la$ may be different
for each shock). A third model is the ``piston" (P) model,
where the ejecta provides pressure but the reverse shock  is assumed to
be an inefficient radiator (either due to low fields or poor acceleration),
while turbulent field growth and acceleration are efficient in the blast wave,
which contributes all the radiation.

If a shock randomizes a large fraction of the total bulk kinetic energy, which
is  comparable to the initial burst energy $E_o$, the comoving synchrotron
luminosity will be  given by $L_{sy}\sim (E_o / 4\pi\theta^2\eta^3) e_{sy}
t_{ex}^{-1}$, where $e_{sy}$ is the synchrotron efficiency relative to the
other mechanisms (see eq.[2.3.4]). Here we used the fact that in the coasting
expansion the expanding shell consists, in effect, of $\sim 4\pi
\theta^2\eta^2$
causally unconnected regions, each subtending an angle
$\sim \eta^{-1}$ at the fireball's centre, and each of comoving dimension
$\Delta R \sim r_d /\eta$;   the comoving spectral intensity is
$I_\nu \sim (L_{sy}/\nu\Delta R^2)\sim (2\nu^2/c^2)\gamma_{sy}m_ec^2$ at the
frequency where self-absorption sets in.  The electron Lorentz factor to be
used
here is $\ga_{sy}={\hbox{max}}[\ga_m , \ga_{ab}]$, where $\ga_m$ is the minimum
Lorentz factor of the electron power law produced in the shock, and
$\ga_{ab}\sim 10^{-3}B^{-1/2}\nu_{ab}^{1/2}$ is the Lorentz factor of electrons
that would be radiating photons at a synchrotron frequency equal to the
self-absorption frequency $\nu_{ab}$. The comoving self-absorption frequency is
therefore
$$
\nu_{ab} =\bigl({ L_{sy} c^2 \eta^2 \over 2 r_d^2 \ga_{sy} m_e c^2
}\bigr)^{1/3}
$$ $$
\sim \cases{ 7.4\times 10^{12}\n^{1/3}\eth^{2/3}(e_{sy}/\ga_m)^{1/3},&
                                                       if $\ga_m >\ga_{ab}$;\cr
             1.1\times 10^{11}\n^{2/7}\eth^{4/7}B^{1/7}(e_{sy}/10^{-3})^{2/7},&
                                                       if $\ga_m <
\ga_{ab}$.\cr}
                                                                  \eqno(2.2.2)
$$
The value of $e_{sy}$ and $B$ to be used here is different for the frozen-in
and
turbulent cases (e.g. eq.[2.2.1]). The minimum electron Lorentz factor $\ga_m$
of the power law distribution behind a shock depends on the bulk Lorentz factor
$\Gamma$, and it may exceed $\Gamma$ by a factor up to $m_p/m_e\sim 10^3$.
Since the reverse shock eventually becomes marginally relativistic,
$(\Gamma_r -1) \sim 1$, while the blast wave (forward) shock achieves a
saturation
$\Gamma_b\sim \eta$  (when $\eta < \Gamma_m\sim 3.3\times
10^5\E^{1/3}r_6^{-2/3}$), we will parametrize the minimum electron Lorentz
factor by
$$
\ga_m = \kappa\Gamma \sim \cases{ \kappa , & (reverse shock);\cr
                             \kappa\eta  , & (blast wave),\cr}
\eqno(2.2.3)
$$
where $1 \siml \kappa \siml 10^3\sim m_p/m_e$. The
lowest energy electrons emit synchrotron
photons with characteristic energy
$\nu_{sy,m}\sim 10^6 B \ga_m^2 \sim 10^6 B \kappa^2 \Gamma^2$ ; in the case
$\ka\sim 10^3\ka_3$, these
are often above the self-absorption values (2.2.2).

The comoving synchrotron cooling time behind a shock characterized by
$\kas\Gas$ will be
$$
t_{sy}\sim { 3.3\times 10^7 \over \ka_{sy}\Ga_{sy} u_B } \sim \cases{
3.1\times 10^3\n^{-1}\la^{-1}\kas^{-1}\Gas^{-1}\eth^{-2}\s,&(\equip);\cr
3.1\times
10^9\n^{-4/3}\E^{1/3}\xi^{-1}\kas^{-1}\Gas^{-1}\eth^{-4}\s,&(\froz),\cr}
\eqno(2.2.4)
$$
which, compared to the comoving expansion time (2.1.2), can be short even for
$\ka\sim 1$ in the equipartition case, but is usually long even for $\ka\sim
10^3$ in the \froz case. The synchrotron energy density can be estimated as
$ u_{sy}\sim n_e(4/3)\sigma_T c u_B \gamma_m^2 t_{ex}$, where the optically
thin free-flight time across the region $\Delta R$ is essentially the
comoving expansion time $t_{ex}$ (2.1.2), and the electron density $n_e$
behind {\it either} one of the shocks is $n_e\sim 10^6 \n\eth^2\Gas^{-1}$
(since it is $10^6\n\eth^2$ in the ejecta or reverse shock, and $10^3\n\eth$ in
the shocked external gas in the blast wave, which has $\Ga=\Gas=\eta$). Thus,
while $\usy$ differs in the two field cases,
$$
\usy \sim \cases{
2\times
10^{-1}\n^{5/3}\E^{1/3}\th^{-2/3}\la\kas^2\Gas\eth^{7/3}\ende,&(\equip);\cr
2.1\times 10^{-7}\n^2\th^{-2/3}\xi\kas^2\Gas\eth^{13/3}\ende,& (\froz),\cr }
\eqno(2.2.5)
$$
the ratio
$$
{\usy \over \ub}= 3\times
10^{-5}\n^{2/3}\E^{1/3}\th^{-2/3}\kas^2\Gas\eth^{1/3}~,
\eqno(2.2.6)
$$
is the same in both.

The comoving IC cooling timescale is given by $\tic=3\mc2 /
(4\sigma_t c \usy\kai\Gai )$, or
$$
\tic=10^{12}\n^{-2/3}\E^{-1/3}\th^{2/3}\kas^{-2}\kai^{-1}\Gas^{-1}\Gai^{-1}
\ub^{-1}\eth^{-1/3}
$$ $$
=\cases{
1.5\times 10^8\n^{-5/3}\E^{-1/3}\th^{2/3}\la^{-1}\kas^{-2}\kai^{-2}
\Gas^{-1}\Gai^{-1}\eth^{-7/3} \s,& (\equip);\cr
1.4\times 10^{14}\n^{-2}\th^{2/3}\xi^{-1}\kas^{-2}\kai^{-1}\Gas^{-1}\Gai^{-1}
\eth^{-13/3}\s , & (\froz).\cr} \eqno(2.2.7)
$$
This includes the case when one of the shocks (with $\kai,~\Gai$) IC-scatters
synchrotron photons from the same shock or the other shock with $\kas,~\Gas$
(i.e., if \sync and IC occur whithin the same shock then $\kas=\kai$,
$\Gas=\Gai$,
or \sync and IC-scattering occur in different shocks then these quantities
differ).
The IC cooling timescales are,
especially for the equipartition case, shorter than the comoving expansion time
(2.2.2) whenever $\kas$ or $\kai$ are large (e.g. $\kappa \sim \mpe\sim
10^3$), and even more so in the blast wave, where also $\Gas\sim\Gai\sim
10^3\eth$. The ratio of the inverse Compton to the synchrotron time is
$$
{\tic \over \tsy} = 4.8\times 10^4\n^{-2/3}\E^{-1/3}\th^{2/3}\kas^{-1}\kai^{-1}
\Gai^{-1}\eth^{-1/3}~,\eqno(2.2.8)
$$
for both the equipartition and \froz cases; using ``cosmological'' parameters,
this will be short for $\kas\kai\Gai \simg 10^5$.

In the case of the blast wave, one must consider also the
combined IC cooling timescale due to scattering the synchrotron photons that
arise in the blast wave plus those that arise in the reverse shock and
travel through the blast wave towards the observer.
(Note that the bulk motion of the radiating material is only mildly
relativistic with respect to the contact-discontinuity, so it is a
reasonable approximation to suppose that $\sim {1 \over 2}$ the radiation
crosses the contact discontinuity).
This timescale is
given by $t_{ic,c}=3\mc2/[4\sigma_T c\kai\Gai ( u_{sy,r}+u_{sy,b})]$,
where $u_{sy,r},~u_{sy,b}$ are the synchrotron photon energy densities due
to the reverse shock and blast wave respectively. Since the energy loss rates
are additive, this combined IC timescale can be obtained from the
appropriate combination of the timescale (2.2.8),
$$
t_{ic,c}^{-1}=t_{ic,rb}^{-1}+t_{ic,bb}^{-1} ~,\eqno(2.2.9)
$$
where rb and bb stand for synchrotron/IC due to reverse/blast and
blast/blast, respectively. Depending on the basic shock model being assumed,
some of these components may be absent, e.g. in the piston model there are
no contributions from the reverse shock, and no combined IC scattering.
\bsk\gbr
2.3 ~ Spectral Components and Fluences
\bsk
The spectrum observed will be made up of a combination of one or two
synchrotron spectra, and one to three IC-scattered spectra (we consider here
only the case where higher order IC scattering occurs in the Klein-Nishina
regime, so its contribution can be neglected relative to the others). The \sync
spectra will be self-absorbed below $\nu_{ab}$ given by (2.2.2), in which case
if $\ga_{ab}>\ga_m\sim \ka\Ga$ the \sync spectrum from  the
blast wave peaks at that frequency; the spectra have
an energy slope $I_\nu\sim \nu^{s_a}$ above that, where
$s_a=-(p-1)/2$ and $p$ is the electron power law energy index above $\ga_m$
defined by $N(\ga)\propto \ga^{-p}$. However, in the case when $\ka\Ga \gg 1$
so $\ga_m \geq \ga_{ab}$, the negative slope $s_a=-(p-1)/2$ starts at
$\nu_{sy,m}\sim 10^6 B\ga_m^2 > \nu_{ab}$. The typical values for the
non-thermal charged particle power  index achieved in shock acceleration are
$p\sim 2-4$, and taking an average  value $p\sim 3$ leads to a \sync slope
above the break $s_a\sim -1$. For a single field value throughout the shock
emission region, the slope below this break will be $+1/3$ while the spectrum
is optically thin, and $+5/2$ below $\nu_{ab}$. In practice, however, it is
unlikely that the field is uniform, and based on the analogy of compact
radio sources, which show an energy slope $\sim 0$ below the break, one might
expect also here an energy slope $s_b$ below the break flatter than either +1/3
or +2.5. In the absence of absorption, we
shall take a fiducial value of $s_b=+1/3$. The corresponding slope of the power
per decade  spectrum $\nu I_\nu$ will be $P_\nu\propto \nu^{\alpha_i}\propto
\nu^{s_i+1}$ where $i=(b,a)$ for frequencies (below, above) the break. This
corresponds to the ``fiducial" values of the power slope $\alpha_b\sim +4/3$
below the break and $\alpha_a\sim 0$ above the break, close to the average
values 1 and 0 (e.g. Schaefer,  \etal, 1992) which have in the past been taken
as a guideline. However, one must be  aware of the fact that there is a
considerable spread about these values (e.g., Band,  \etal, 1993).

The lab-frame frequency of the \sync break or turnover will be given
either by the comoving self-absorption frequency (2.2.2), or (for the larger
$\kas\Gas$ in either shock) by $\nu_{sy,m}\sim 10^6 B\ga_m^2$,
in both cases blueshifted  by $\eta=10^3\eth$ (the bulk Lorentz factor of
both the reverse shock and the blast wave, in the observer frame). In the
latter case, the lab-frame \sync turnover frequency is
$$
\nu_\syt\simeq \cases{
4\times 10^{11}\n^{1/2}\la^{1/2}\kas^2\Ga_{sy}^2\eth^2 \Hz,& (\equip);\cr
4\times 10^{8}\n^{2/3}\E^{-1/6}r_6^{1/2}\xi^{1/2}\kas^2\Ga_{sy}^2\eth^3 \Hz ,
                                               & (\froz) ,\cr} \eqno(2.3.1)
$$

The IC-scattered spectrum will show a corresponding IC break or turnover
at a lab-frame frequency $\nu_\ict \sim (4/3)\kai^2\Gai^2\nu_\syt \sim
1.3\times 10^9 B \kas^2\kai^2\Gas^2\Gai^2\eth$, or
$$
\nu_\ict \simeq\cases{
5.3\times
10^{11}\n^{1/2}\la^{1/2}\kas^2\kai^2\Gas^2\Gai^2\eth^2\Hz,&(\equip);\cr
5.3\times 10^8
\n^{2/3}\E^{-1/6}r_6^{1/2}\xi^{1/2}\kas^2\kai^2\Gas^2\Gai^2\eth^3
                                        \Hz, &(\froz).\cr} \eqno(2.3.2)
$$
The IC-continuum slope below the turnover will be the same as for the
synchrotron spectrum, and that above the turnover will also be the same as long
as the electron slope of the IC-scattering electrons $p_{ic}$ is even slightly
steeper than that of the of the synchrotron electrons $p_{sy}$, in the case
of scattering of reverse-shock
photons on blast wave electrons. If the synchrotron and
IC electron slopes are the same, the IC slope above the turnover is
logarithmically flatter than that of the synchrotron spectrum. This logarithmic
flattening, if present, will be ignored below, since it does not affect
substantially the discussion. Alternatively, if the synchrotron electrons have
a slope $p_{sy}$ steeper than $p_{ic}$, the IC-scattered photon energy spectrum
has a slope of $-(p_{ic}-1)/2$, which we will take for numerical examples to be
of order $-1$ (corresponding to a power per decade or fluence per decade slope
$\alpha_a=0$).

The total spectrum will then show a number of spectral components due to the
various \sync and IC combinations possible between the two shocks, compatible
with the assumptions made about the field origin. In the simpler case of
frozen-in magnetic fields in the ejecta, which will affect only the reverse
shock (and assuming that turbulent magnetic field generation in both shocks
is inefficient, $\la \ll 1$), we have three components: one \sync from the
reverse shock, and two IC, one from the reverse and one from the blast wave.
These may be labeled $(sy,r),~(ic,rr),~(ic,rb)$.
In the case where turbulent field generation is efficient enough to achieve
a non-negligible fraction of the equipartition magnetic field value in both
shocks (i.e., a field larger than whatever frozen-in component that may be
present) there are five spectral components: two \sync, one from each shock,
and three IC. These may be labelled $(sy,r),~(sy,b),~(ic,rr),~(ic,bb),
(ic,rb)$.
Intermediate situations may exist if turbulent field generation is efficient
in only one shock but not the other, which can be treated in a similar manner.

Each spectral component will produce a total photon energy fluence
$S~(\erg\cmsqi)$, which is some fraction $\siml 1$ of the maximum
bolometric fluence available from the particular shock in question. In
the most common case when $\eta <\Gamma_m=3.3\times 10^5\E^{1/3}r_6^{-2/3}$,
most of the initial fireball energy is carried as kinetic energy of motion
of the ejecta, which is completely re-randomized in the blast wave and the
reverse shock. We shall therefore assume that the maximum bolometric fluence
possible $S_o=S_r+S_b$ is divided about equally between the reverse shock
and the blast wave, and is
$$
S_o= (\varep_r + \varep_b)(S_o/2) =
E_o/4\pi\theta^2 D_L^2 =10^{-6} \E \theta^{-2}D_{28}^{-2} \erg\cmsqi~,
\eqno(2.3.3)
$$
where $\varep_r\sim 1,~\varep_b\sim 1$ are parameters describing what fraction
of the total bolometric fluence is carried by the reverse and blast wave
shocks,
and $D_L$ is the luminosity distance, $D_L=(2c/H_o)[(1+z)-(1+z)^{1/2}]
\sim 10^{28} h^{-1}$ for $z\sim 1$ in an Einstein-de Sitter universe, or
its appropriate Newtonian scaling for galactic models.

For each shock we can define the \sync and IC radiative efficiencies $e_{sy},~
e_{ic}$ for individual spectral components as
$$
e_{ic}={\tic^{-1} \over \tic^{-1}+\tsy^{-1}+\tex^{-1} }~~,~~
e_{sy}={\tsy^{-1} \over \tic^{-1}+\tsy^{-1}+\tex^{-1}  }~, \eqno(2.3.4)
$$
where the energy-loss timescales have to be evaluated for the particular
shock $j$ and the particular spectral component (combination of shocks $j,k$),
e.g. in the IC case, one could have $t_{ic,rr}$, $t_{ic,bb}$ or $t_{ic,rb}$.
The corresponding photon energy fluences from each shock or shock combination
due to each mechanism (giving the different spectral component) are
$$
S_{ic,jk}= \varep_j (S_o/2) e_{ic,jk}~~,~~ S_{sy,j}= \varep_j (S_o/2)
e_{sy,j}~,
\eqno(2.3.5)
$$
each extending over an energy range that goes significantly below and above
the turnover (break), to an extent depending on the upper and lower limits of
the electron energies.
\bsk\gbr
2.4~~ Gamma-ray, X-ray and Optical Fluences.
\bsk
We consider now more specifically the fluences in three frequency bands
$\Delta\nu/ \nu\sim 1$ centered around $1.2\times 10^{20}\Hz$ ($\gamma$-rays
near 0.5 MeV), $3\times 10^{17}\Hz$ (medium/soft X-rays) and $10^{15}\Hz$
(O/UV), matching approximately the HETE instrument bands (Ricker, 1992). We
take
the synchrotron spectrum to be given by a broken power law in energy flux,
$I_\nu \sim \nu^{s_i}$, where $s_i=[s_b,~s_a]$ for $\nu \ltgt \nu_{syt}$ ,
in terms of the observed lab-frame $\nu_{syt}$ of eq.
(2.3.1-2.3.2). The IC-scattered spectrum will also be a broken power
law with the same energy indices, $I_\nu\propto \nu^{s_i}$, while the
corresponding
power per decade (and fluence per decade) spectra have slopes $P_\nu=\nu I_\nu
\propto \nu^{\alpha_i}$ where $\alpha_i=s_i+1$.  In a double-logarithmic plot
the IC spectrum is shifted with respect to the synchrotron spectrum upwards and
to
the right by a factor $(4/3)\kappa^2\Gamma^2$. (We assumed, for simplicity,
that
the scattering electron energy index $p$ satisfies $s_a \siml (p-1)/2$,
although
this is not necessary). The $\gamma$-ray fluence in the $\nu\sim \nu_\ga \simeq
0.5 \MeV$ band will then be given by the sum (or in practice by the largest) of
the various IC and synchrotron spectral contributions in that band. If
self-absorrption effects can be ignored, this is
$$
S_\ga = \sum_{n,jk} A_{n,jk} S_{n,jk} \left( \nu_\ga
/\nu_{nt,jk}\right)^{\alpha_i}\flu
\eqno(2.4.1)
$$
where $n$ stands for \sync or IC, $nt$ stands for the turnover $syt$ or $ict$,
$jk$ stands for the shock combinations $rr,~ bb$ or $rb$ (where relevant), the
constant $A_{n,jk} \sim $ 1/few is taken for numerical estimates to be $A=0.2$,
and $\nu_{nt,jk}$ stands for the n-mechanism turnover frequency of the shock
combination $jk$.
Similarly the X-ray fluence in the $\nu\sim \nu_x \simeq
3\keV$ band is given by
$$
S_x = \sum_{n,jk} A_{n,jk} S_{n,jk} \left( \nu_x
/\nu_{nt,jk}\right)^{\alpha_i}\flu
\eqno(2.4.2)
$$
and the O/UV fluence in the $\nu\sim \nu_u\simeq 10^{15}\Hz$ band is given by
$$
S_u = \sum_{n,jk} A_{n,jk} S_{n,jk} \left( \nu_u
/\nu_{nt,jk}\right)^{\alpha_i}\flu
\eqno(2.4.3)
$$
The above are without self-absorption, and generally we take $\alpha_b,\alpha_a
=
4/3, 0$, although observationally other values are also expected. However, in
many cases the spectra are more complicated than (2.4.1-2.4.3) because
self-absorption becomes important at the lower energies. For $\nu_{ab} <
\nu_m$,
the slope is $\alpha_a$ for $\nu >\nu_m$, it is $\alpha_b~(\sim 4/3)$ for
$\nu_{ab}
<\nu <\nu_m$, and 3 for $\nu <\nu_{ab}$ (e.g. 0, 4/3, 3 for decreasing
energies).
For $\nu_{ab} > \nu_m$, it is $\alpha_a$ for $\nu >\nu_{ab}$, it is 3.5 for
$\nu_m
<\nu <\nu_{ab}$, and 3 for $\nu <\nu_m$ (e.g. 0, 3.5, 3 for decreasing
energies).
In these cases, an individual synchrotron component has two breaks,
and the corresponding IC-scattered component will also reflect this fact.
Notice that the power in a certain band (e.g.
eqs. 2.4.1-2.4.3) depends on the slopes $\alpha_i$, but the power at  the break
energies $S_{n,jk}$ (eq. 2.3.5) is independent of these slopes. Thus, unless
the bands are very far from the break energies and the slopes are very
different from those assumed here, the band fluences should not be  too
different from those estimated with the fiducial slopes 4/3, 0. The limiting
$\ga$-ray sensitivity of BATSE (e.g. Fishman, 1992) is $S_\ga\sim 10^{-8}\flu$,
while for bursts of $10\s$ that of HETE (Ricker, 1992)  is $S_\ga\sim 3\times
10^{-7}\flu$, and in the X-ray and UV bands the sensitivity  limits are
$S_x\sim 8\times 10^{-8}\flu$ and $S_u\sim 10^{-8}\flu$.

The calculated positions of the spectral breaks, and the slopes of the
power-law segments of the spectra, are shown in the figures. In these figures,
the breaks are depicted as being sharp. In reality they will be smoother, not
only because an individual electron emits broad-band radiation but also because
the observed flux comes from different parts of the fireball whose doppler
shifts have a spread of order 2 (from $\sim \eta$ to almost $2\eta$).

\bsk\gbr
\ctl{\bf 3.~~ Results}
\bsk
3.1.1~~ General Features
\bsk
In general the fluences scale as $D^{-2}$, and as a positive power (usually not
unity,
due to the efficiency factors) of the intrinsic total initial energy $E_o$.
They also
scale as a negative power of the beaming half-angle $\theta$, which is
generally not
$\theta^{-2}$ (again due to the $\theta$ dependence of the efficiency and
turnover
factors). The fluences also scale as a positive power of the external density
$\n$,
generally between $1/3$ and $2/3$.
They also depend on $\eta$; it is the burst timescale, however, which is the
most
sensitive to this parameter.

In the cosmological case we adopt $E_o =10^{51}\E\ergs$ and
$D=10^{28}D_{28}\cm$. For $\n=1$ the burst duration for $\eth=1$ is $t_L=5\s$
and the maximum photon bolometric fluence is $S_m\sim 10^{-6} \E
D^{-2}\theta^{-2} e_t$, where $e_t$ is the total radiative efficiency which
depends (usually nonlinearly) on $\E,\n$, etc. For smaller $\eta$ and/or
smaller $\n$ the observed timescale gets longer, and vice versa.

In the galactic halo case we use, e.g., $E_o=10^{41}\ergs,~D=1.5\times
10^{23}\cm=50\Kpc$, while for the galactic disk we use generally
$E_o=10^{39}\ergs$, $D=3\times 10^{21}\cm= 1\Kpc$. The maximum bolometric
fluences $S_m$ are (in principle) similar to the  cosmological case, for this
choice. However, especially in the frozen-in field models,  which are
inefficient radiators, higher $E_o$ are occasionally needed to reproduce the
observed $\ga$-ray band fluences. Compared to the cosmological case, the
durations are  comparable for the halo (where typically $\n=10^{-3}$), but for
the galactic disk  (where typically $\n=1$) they are shorter (see eq.2.1.4).
The typical external densities  are the interstellar values expected for the
halo/disk case, unless  the burst occurs inside a denser pocket of material,
e.g. a wind or nebula that preceded the event. For the lowest interstellar
densities, acceptable $\ga$-ray fluences require  generally large $\ka\sim
10^3$, while for higher densities $\n\simg 1$ $\ka\simg 40$ is  required.

The spectrum depends on the value of $\ka$ and $\la$ in the shocks, and we
generally
assume that $\ka$ and $\la$ are the same (or do not differ much) for both
shocks.
We shall consider values of $10^{-6}\siml (\la,\xi) \siml 1$, $10^{-3}\siml \n
\siml
10^3$, $\ka=~1,~40,~10^3$, $1\siml \eta \siml 10^4$ and slopes
$\alpha_i=4/3,~0$ in
most of the illustrative cases discussed below.
\bsk
3.1.2~~ Frozen-in Field Models (F)
\bsk
In this model magnetic fields are assumed to have built up in the ejecta at the
beginning of the fireball's history to some fraction $\xi$ of the equipartition
value with the total initial disposable energy $E_o$. Magnetic fields in
the blast wave and reverse shock due to turbulence, etc., are assumed
to be weak, so that \sync radiation is important only in the reverse shock,
where the frozen-in field is characterized by $\xi$, but IC scattering in
the blast wave is important, because it efficiently accelerates electrons
to high $\gamma$. This scenario has
only three radiation components. It is also the least efficient, the \sync and
IC  radiative efficiencies being generally smaller for $\xi=1$ than for the
$\la=1$  shock-turbulent or piston models with the same external density
(which we discuss below). This
is because $\xi=1$ means equipartition at the original explosion, but
then the field energy decreases adiabatically with the expansion;
even after compression (by a modest factor) in the reverse shock, it is well
below
equipartition with the bulk kinetic energy (see eq. 2.2.1).

At cosmological distances, if one uses the low value $\ka=1$ in both shocks,
and
$\theta=\xi=1$, the fluences in all three wavebands are below the detection
threshold
for $D_{28}=1$, and are of course smaller for $\xi <1$. However the situation
is
better for the case when $\kappa\gg 1$, especially if the burst is beamed.  For
$\theta=10^{-1}$, $\kappa=10^3$ and $\n=\xi=1$, the $\ga$-ray fluences are
well above threshold, and the major break is in the neighborhood of the
$\ga$-ray
band, so that the X-ray fluence is significantly lower, for a range of burst
durations
comparable to that in observed GRBs. A spectrum for this frozen cosmological
(FC)
model is shown in Fig. 1a, where the individual components are also indicated.
The frozen-in models are easily detectable up to $D_{28}\siml 1$ for values of
$\xi$ not too far below unity. For the same parameters but $\xi\siml 10^{-2}$
the
$\ga$-ray fluences would drop below detectability, unless $D_{28}\siml 10^{-1}$
or
the external medium is denser. For instance, $\xi=10^{-3},~\n=10^3,~D_{28}~=1$
yields
large $\ga$-ray fluences and reasonable X-ray fluences.

At galactic halo and galactic disk distances, because of the lower total
energy,
the deceleration radii $r_d$ are generally smaller. A burst of given duration
requires a smaller $\eta$ (for a given $n_{ext}$). The magnetic fields
are weaker, and the cooling times are longer while the
comoving expansion time is shorter. Therefore the efficiencies are lower  than
in the cosmological case, and their relative values change. Thus, in the
galactic cases the synchrotron efficiency generally exceeds the  IC efficiency,
even for large values of $\kappa\sim 10^3$ (unless the burst  occurs inside
very dense clouds $\n \simg 10^6\cmcui$). The spectra are consequently
different from those of the same model at cosmological distances.  Because of
the intrinsically lower radiative efficiency of the frozen-in models  (compared
to the turbulent and piston models), detectable fluences are obtained  only for
relatively large (by galactic standards) total burst energies, e.g.
$E_o=10^{45}\ergs$ and higher for the halo or $E_o=10^{43}\ergs$ for the disk.
Acceptable $\ga$-ray  fluences are obtained mainly for beamed, high $\ka$
cases, e.g.
$\th=10^{-1},~\ka= 10^3$ and $\xi=1$. A spectrum for the frozen halo (FH) case
is
shown in Fig. 2a, for these parameters, $\n=10^{-3}$ and two values of $\eta$
corresponding to 12 s and 0.5 s durations. A  spectrum for the frozen galactic
disk
(FD) case ($\n=1$) is given in Fig. 3a, for 5 s and 0.25 s.  In both halo and
disk
cases the major break is below all three observational bands,  so the fluences
at UV, X-rays and $\ga$-rays are all comparable, $S_x/S_\ga\sim  S_u/S_\ga \sim
1$.

The general behavior of the spectra for bursts of various durations is shown,
for the frozen-in cosmological (FC) model in Fig. 4, for the same parameters as
Fig. 1a but variable duration $t_L$. The three break energies tend to be
softer for the longer total durations, and also the fluence levels within the
same
plateau  are smaller  for longer durations. However, for a fixed energy band
the fluence can also increase  rather steeply as the duration is taken longer,
if one of the breaks moves across that energy band.
\bsk\gbr
3.1.3~~ Turbulent Field Growth Models (T)
\bsk
In this model turbulent magnetic fields are assumed to build up to some
fraction  $\la$ of the equipartition value in both the reverse and blast wave
shocks (not necessarily the same in both, although here we assumed them to be
equal). As a result there are  generally five different spectral components,
although often one component dominates over several of the bands $\ga,x,u$.
For values of $\kappa\sim 1$ in both shocks, i.e. $\ga_m\sim \Gamma$,
the  \sync efficiency is generally larger than that for the IC process. For
larger $\ka$ the importance of IC is larger, and in the cosmological case
(where $E_o$ is larger so $r_d$ is larger and $B_d$ smaller) it dominates the
radiative efficiency. At galactic distances, $E_o$ and $B_d$ are smaller and
synchrotron remains mostly dominant, even when IC is important from the
spectral point of view.

For cosmological distances, observable $\ga$-ray fluences are obtained for
$\ka=1$ if $\la\simg 10^{-3}$, and for $\ka \sim 10^3$ even for values $\la $
well below $10^{-6}$. However, for $\la \gg 10^{-6}$ this model is also
extremely  efficient in producing radiation at other frequencies below the
$\ga$-ray
band. A turbulent cosmological (TC) spectrum for standard parameters $\n=1$,
$\kappa=10^3$, $\la=10^{-6}$, $\th=10^{-1}$, $\eth=1$, $t_L=5\s$ is given in
Fig. 1b,
showing the  five spectral components. It is seen that, even for this low field
case
($10^{-6}$ of the equipartition value) the $\ga$-ray fluence is high. This is
true also for higher $\la$ values, but the X-ray to
$\ga$-ray fluence ratio $S_x/S_\ga$  is then closer to unity, whereas for the
lower values such as $\la=10^{-6}$ this ratio is of  order $\siml 0.03$.  It is
worth noting that the spectrum for the frozen-in model with $\xi=1$ plotted in
Fig. 1a is essentially the same as that for the ``turbulent''
model at the same distance with $\la=10^{-6}$ (in Fig. 1b).
The reason for this is that for this choice of $\la$ and $\xi$ the fields in
the shock at deceleration have the same magnitude (eqs. 2.2.1), so the dominant
reverse-shock \sync, reverse-shock IC and blast wave combined IC are the same,
while
the blast wave \sync and blast wave IC spectra are below the other three
components.

For galactic halo and disk distances the turbulent model $\ga$-ray fluence is
generally
high, even for densities somewhat lower than the standard interstellar values
($\n=
10^{-3}$ halo, $\n=1$ disk), especially at the higher $\ka$ values. As in the
galactic
frozen-in models, due to the lower field value the breaks occur at relatively
low
energies and the flat (slope 0) spectrum extends to energies lower than in the
cosmological case. As a result, there is an increased tendency to produce a
relatively
large X-ray fluence, both the $\ga$- and X-ray bands falling above the main
break.
There are some exceptions to this situation, e.g. in the disk case for
$\th=10^{-1},~\la=10^{-3},~\ka=10^3$ the X-ray fluence is well
below the $\ga$-ray fluence for short bursts $\eta=10^2$, $t_L=0.2\s$, but they
are
comparable for the longer bursts $t_L\simg 1\s$, and for longer bursts the UV
fluence
can also be significant. The spectra characteristic of such short and long
duration
halo and galactic disk turbulent models are shown in Figs. 2b and 3b.

The spectra of bursts as a function of burst duration is shown for the
turbulent
cosmological model TC in Fig. 5, for fields in the shock which are
$\la=10^{-6}$
below equipartition. These spectra show the same break softening and plateau
fluence decrease with increasing duration as seen in the frozen-in case, but
less pronounced. The reason for the  difference is the different $\eta$
dependence (or $t_L$ dependence) of the field $B$ at the shock (eq. 2.2.1). For
this choice of $\xi=1$ (frozen) and $\la=10^{-6}$ (turbulent) the fields (and
spectra) are the same at $t_L=5\s$ ($\eta=10^3$), but differ  at other
durations (other $\eta$). For this low $\la$, the two additional spectral
components in the turbulent model just begin to become noticeable at the
longest durations. The two lower energy breaks are due to the reverse shock
\sync and IC, while the two highest are the blast wave self-IC and combined IC
(the blast wave \sync is just visible around $10^3\s$ near $10^{17}\Hz$). The
MeV band is in the neighborhood of the reverse IC component (redshift is not
included in the cosmological plots shown).

In Fig. 6 we show the turbulent halo model spectra as a function of duration.
In this case the two main breaks are the reverse and blast \sync breaks, the
MeV band  falling near the blast \sync break. The turbulent galactic disk model
is similar, but  the MeV band is farther up on the plateau above the blast
\sync
break, while the reverse  \sync is closer in fluence and energy to the blast
\sync component.
\bsk
3.1.4~~ Piston Models (P)
\bsk
The physics of these models is the same as in the turbulent field growth cases
discussed before, except that there are no radiative contributions from the
reverse
shock, only from the  blast wave. The spectra have only two components, \sync
and IC.
The relative importance of the blast \sync and IC contributions are changed,
relative
to the two-shock turbulent growth model, because the much more abundant reverse
shock photons which previously acted as  seeds for the blast wave IC are now
absent. For this reason, the blast wave \sync spectrum  dominates in almost all
cases. The spectrum being so simple, the value of the fluence at each band is
determined simply by the slopes assumed and by whether the observed band is
above or below the \sync break. It is therefore strongly dependent on the field
strength $\la$ and the particle minimum energy $\ka$.

The cosmological piston (PC) model has a high bolometric radiative efficiency
for a
wide range of $\la$ and $\ka$.  The highest $\ga$-ray fluences (for a slope 0
above the break) are obtained when the break is below the $\ga$-band, which
requires $\la$ and/or $\ka$ not to be too large. Values of $\ka\sim 10^3$ are
too large, but high $\ga$-fluences are obtained for $\ka\siml 10^2$, for a
range of burst durations. When $\ka$ is much lower than this, and/or  $\la$ is
very low, the X-ray fluence can become comparable to the $\ga$-ray fluence. The
spectrum of a cosmological piston model and its components is shown in Fig. 1c.

The piston models in the halo and galactic disk (PH, PD) have a lower radiative
efficiency than the cosmological piston case, due to the lower energy density
and
corresponding lower  magnetic field in the blast wave. The total energies
required to obtain acceptable  $\ga$-ray fluences are comparable to those in
the turbulent growth model, $E_o=10^{41} \ergs$ (halo) and $E_o=10^{39}\ergs$
(disk). Due to the lower field the break energy  appears at low frequencies,
typically below the UV band.  High bolometric fluences are  obtained for large
$\ka\sim 10^3$, giving spectra whose fluences in the $\ga$-ray and  X-ray bands
are comparable for a high energy slope of 0 for durations $t_L \simg 1\s$, but
ratios $S_x/S_\ga \siml 0.03$ are obtained in the short bursts ($t_L \siml
1\s$).

In Fig. 7 we show, for a fixed density, the piston cosmological (PC) spectra
expected as a function of the burst duration, for $\la=10^{-3}$. The piston
models have only blast wave spectral components, and in this case the MeV band
falls near the blast \sync break, while the blast IC component is seen at the
highest energies.

In Fig. 8 the piston disk case (PD) is shown, where again the MeV band falls
nearest to the blast \sync break, while the blast IC component is not
sufficiently strong to show up. The corresponding piston halo model shows
similar properties.
\bsk\gbr
3.1.5~~ Gamma-ray and X-ray Fluences
\bsk
The spectra previously discussed refer to 'standard' densities, and it is
interesting to consider other values as well. In the density-duration parameter
space, we have calculated the $\gamma$-ray band $S_\ga$ fluence contour levels
and also the ratio $S_\ga/S_x$, these two bands being centered around 0.5 MeV
and 1 KeV. These are the most likely to be of immediate use in
constraining models. Similar contour levels for the $S_\ga/S_u$ ratio can also
be made but are omitted here for brevity. The $S_\ga$ contours start at the
approximate BATSE threshold of $10^{-8}\flu$ and higher values, while the
$S_\ga/S_x$ contours go from $1$ to $10^2$. Of course, other values also occur,
but these delimit the most interesting range between comparable $\ga$ to
$X$-ray ratio and the usual ``X-ray paucity" value of $S_x/S_\ga \siml$ few
percent. For each model these two sets of contours are shown simultaneously in
Fig. 9. The left column shows the cosmological models (frozen, turbulent and
piston from top to bottom), the middle column the halo models and the right
column the galactic disk models (same distribution). The frozen and turbulent
cosmological $S_\ga$ and $S_\ga/S_x$ contours behave fairly similarly in the
$n,t_L$ plane, not surprisingly since the field values were chosen
approximately similar. The higher fluences are in the $n\simg 1,~t_L\simg 5\s$
upper right quadrant, but detectable  $S_\ga$ is obtained over much of the
range $10^{-1}\siml t_L\siml 10^3,~10^{-3}\siml  n\siml 10^3$. The region where
the $S_\ga/S_x$ ratio is larger than 30 is also in the  upper right quadrant,
so longer bursts are better at satisfying the X-ray paucity constraint. The
cosmological piston contours are much simpler, due to the simpler (essentially
one component) structure: burst in a medium of $n\siml 10^2$ and with $t_L\simg
1\s$ are all above threshold, and those at all densities with $t_L \siml 12\s$
satisfy a gamma/X ratio $\simg 30$. Here it is the short bursts that satisfy
X-ray paucity. This difference between frozen/turbulent and piston is due to
the fact that, for this particular choice of $\la,\xi$ and $\ka$ the former
have the $\ga$ band halfway up the break slope, while the latter has it on the
plateau above the break, and the break softens with increasing duration. The
details of the behavior, it must be stressed, are a function of the choice of
parameters.

For the halo models (middle column of Fig. 9) the frozen models are above $\ga$
threshold everywhere except at long durations and low densities, higher $S_\ga$
occurring for shorter durations and higher densities. For the frozen model the
X-ray  paucity ratio of 30 is violated almost everywhere, while gamma/x ratios
$\simg  10$ are obtained for $t_L\simg 5\s,~n\simg 10^2$. The turbulent and
piston halo are above $\ga$-threshold essentially everywhere, but turbulent
satisfies a gamma/x ratio $\simg 30$ for $n\siml 10^{-2}$ and durations $0.2\s
\siml  t_L\siml 5\s$, while the piston does it for all densities and $t_L\siml
5\s$.

For galactic disk models (right column) the frozen models behave similarly to
the halo case (middle column), but the region where the gamma/x ratio is
larger than 10 is smaller ($\n\simg 10^2,~10^2\siml n \siml 10^3$). The
turbulent disk has a somewhat larger region in the low density, short time
quadrant where this ratio $\simg 10$ is satisfied, and the piston disk
satisfies it for all densities and $t_L\siml 1\s$.

Also shown in Fig. 9 are the lines below which impulsive bursts are not
possible, shown as full straight lines running diagonally from top left to
bottom right in the halo and galactic disk cases. In the cosmological cases
these lines are outside the figures for the range of values used, so this
restriction does not apply. Short bursts  below and to the left of this line
would require, in the impulsive burst limit (energy deposition time $\ll$ burst
duration $t_L$), a value of $\Gamma=\eta$ which exceeds the maximum value
compatible with dynamical requirements, $\eta > \Gamma_m=  3.3\times 10^2
E_{42}^{1/3} r_6^{-2/3}$ (e.g. \Mesz, Laguna and Rees, 1993). Bursts below this
line, therefore, are only consistent with an interpretation where the timescale
$t_L$ is not given by the dynamics as a function of $\eta$, via eq.(2.1.4) (see
also below).
\bsk\gbr
\ctl{\bf 4.~~ Discussion}
\bsk
We have calculated the radiation spectra of gamma-ray burst sources arising in
the  blast wave produced when relativistic ejecta encounter the external
medium.  While there are a number of uncertainties concerning the field
strength and the particle acceleration efficiency in the blast wave and reverse
shocks, a simple  parametrization has been used to determine the range of
physical conditions that can be  encountered. For models where turbulent
magnetic field growth occurs in the shock the  field strength is characterized
by $\la$, which is the ratio of the magnetic to the particle energy in the
shock, while frozen-in fields are characterized by $\xi$, which is the
corresponding ratio at the time of the initial impulsive event. The particle
acceleration is characterized by the ratio $\ka$ of the minimum electron
Lorentz factor in units of the bulk Lorentz factor of the burst. The range of
parameters  explored is that which reproduces the observed gamma-ray fluences
and the overall burst  durations, $10^{-8}\flu \siml S_\ga\siml 10^{-4}\flu$
and $10^{-1}\s \siml t_L\siml 10^3\s$. Temporal substructure on shorter
timescales is expected, since the external medium is likely to be lumpy and the
shocks are expected to be  unstable. These gross features of GRBs are easily
explained  by the impulsive energy deposition model with physically reasonable
values of these parameters, within the  context of the models discussed here.

A natural explanation for the non-thermal, power-law spectrum seen in the
bursts in the gamma-ray range is provided by the \sync and inverse Compton
radiation  losses expected from the power law relativistic electrons
accelerated in the shocks.  Also the presence of breaks by about one unit in
the power law finds a natural  explanation in this model. These are associated
with the minimum energy to which relativistic electrons are accelerated.
Steeper breaks can also arise from \sync  self-absorption, which we have
evaluated here for the case of relativistically expanding  sources. This is
important in some cases, but it appears usually below the UV  range.

The breaks do not appear at a preferred energy, the latter depending on the
parameter $\eta$ and external density as well as $\kappa$. For a fixed assumed
fraction of the equipartition energy density $\la$, the magnetic field at the
shocks depends on $\eta$  and the break energies depend on $B$ and powers of
$\Gamma$. Since for the reverse shock  $\Gamma_r$ is always of order unity,
the blast wave $\Gamma_b=\eta$, the $\eta$  (or $t_L$) dependence of the break
energy is stronger for the blast wave components than  for the reverse
components. The softening of the `main'  piston spectral break (due to the
blast wave) with decreasing $\eta$ (increasing  baryon loading, or increasing
burst duration $t_L$) is more pronounced than the softening of the `main' break
(due to the reverse shock) in the turbulent and frozen-in  models, but it is
comparable to the softening of the third, combined IC break (due to the blast
wave) in these models.

Specific constraints are obtained by comparing the fluence in different parts
of the spectrum. In particular,  the usually invoked X-ray paucity constraint
$S_x/S_\ga \siml 0.03$ (e.g. Laros, \etal, 1984), if applied universally,
provides
stringent restrictions on the type of models which are acceptable. Only the
cosmological models are able to satisfy this condition over the entire range of
durations $120^{-1}\s\siml t_L\siml 10^3\s$: turbulent cosmological models are
satisfactory for relatively low values of $\la\siml 10^{-6}$, piston
cosmological
models are satisfactory over a wide range of values of $\la$, and frozen-in
cosmological models require relatively high values of $\xi \simg 10^{-1}$. On
the
other hand, the galactic models produce lower magnetic fields in the shocks,
and a
direct consequence of this is that their breaks occur generally at energies
below
the X-ray band. This holds for impulsive bursts, where $t_L$ is determined by
$\eta
<\Gamma_m= 3.3\times 10^5 E_{51}^{1/3} r_6^{-2/3}$ (eq.2.1.4), for durations
$t_L
\simg 1\s$. Galactic models would therefore not produce $\ga$-ray breaks, and
would
also predict rather high $S_x/S_\ga \sim L_x/L_\ga \sim 1$ fluence ratios
violating
the X-ray paucity ratio as defined above, except for very short bursts with
(impulsive) duration $t_L\siml 1\s$ given by eq.(2.1.4). A possible way around
this
might be if the bursts are not impulsive, i.e. the original energy input has an
intrinsic, longer timescale which is identified as the burst duration, e.g.
it is $t_w$ for a wind model, or $t_{accr}$ for an accretion model. In Fig. 9,
impulsive models can only exist above and to the right of the diagonal solid
line (in the cosmological models, this line is outside the figure boundaries).
Below this line, in halo and galactic disk models, only non-impulsive models
are allowed, where $\eta$ and the duration are independendent parameters not
connected by eq.(2.1.4). The duration could, in this case, be given, e.g., by
the
wind duration $t_w$ discussed in \S2.1, or by an accretion time $t_{acc}$.
For non-impulsive models, in Figs. 1 through 9, the
duration ($t_L$) values plotted would not be the real duration; the ``duration"
values in Figs. 4-9 can, however, be simply re-interpreted as a quantity
proportional
to an inverse power of $\eta=E_o/M_oc^2$, a characteristic of the model, while
the
real duration is an extra model parameter. The cosmological models, however,
can
satisfy the impulsive approximation throughout the region of parameter space
here
considered, so for these the duration $t_L$ of eq.(2.1.4) can be
self-consistently
determined from $\eta$. (Non-impulsive models could also be considered in the
cosmological case, at the price of introducing a model-dependent timescale, but
this is not necessary).

There is a well-defined density dependence of the fluences, as seen from Figs
1d, 2d and 3d for the piston model (this holds also for the other two
models). The reason for this is that the deceleration radius (2.1.1) depends on
density, and this determines both the duration (2.1.4) of the burst and the
field
strength (2.2.1) at the shock. While the break energy behaves monotonically
with
$\n$, the  fluence does not, since that depends on the ratio of \sync to IC
efficiencies which is a more complicated function of the density.
The density effect is also seen in Fig. 9, where the
lines of $S_\ga=$ constant and $S_\ga/S_x=$ constant are plotted in the
$n_o,~t_L$ plane. This defines the regions where both an ``X-ray paucity"
constraint and a minimum $\ga$-ray fluence constraint can be satisfied for the
three models considered, in both the galactic and cosmological cases and for
various values of the field generation parameter $\xi$ (or $\la$) and  the
particle acceleration parameter $\ka$. It is seen that, in general, an external
density $n_o\simg 1\cmcui$ is required to produce acceptable fluences. This
would
be an argument against having GRB in a galactic halo. In the case of GRB in
our galactic disk  it would predict that they are seen only inside spiral arms,
where the ISM density is the standard value $n_o\sim 1\cmcui$ or greater (but
the X-ray paucity constraint and burst duration constraint cannot be easily
satisfied, unless additional assumptions are made, as discussed above).
Strong $\gamma$ emission could arise if the electron spectrum were flatter
(with $p<3$), so that synchrotron emission from the highest-energy electrons
was
the dominant process (cf. Begelman \etal 1993).
In the case of cosmological bursts, Figure 9 suggest that the GRB
occur in (distant) spiral galaxies or other ``galactic" environments where the
density is of the order or above the standard ISM density of $1\cmcui$.
However, it is possible that cosmological GRB which have wandered out of their
host galaxy, or galactic GRB in a halo, could be inside a ``local high density
pocket" produced, e.g. by a precursor wind, and might therefore have fluences
comparable to those of bursts in a high density ISM.

The spectral calculations are used also to make predictions of the optical/UV
and  the GeV/TeV fluences, as well as the X-ray fluences expected from bursts
with a given  MeV-band fluence and duration. These fluences can be compared
with the limiting  sensitivities of BATSE (e.g. Fishman, 1992) $S_{\ga,m}\sim
10^{-8} \erg\cmsqi$ in the  $\gamma$-ray band, and of HETE (Ricker, 1992),
$S_{\ga,m}/t_L\sim 3\times 10^{-8} \erg\cmsqi \si$, $S_{x,m}/t_L \sim 8\times
10^{-9}\erg\cmsqi\si$ and $S_{u,m}/t_L \sim 10^{-9}\erg\cmsqi\si$ in the
$\ga$-ray, X-ray and O/UV bands respectively ($10\sigma$ limits). For a typical
burst duration $t_L\sim 10\s$, the HETE fluence sensitivity limits are
$S_\ga\sim
3\times 10^{-7}\flu$, $S_x\sim 8\times 10^{-8}\flu$  and $S_u\sim 10^{-8}\flu$.

Significant X-ray fluences are predicted in some of the models. For a given
(observable) $\ga$-ray fluence, even if one imposes restrictions on the model
parameters to satisfy  an X-ray paucity constraint (e.g., based on current
observations, $S_x/S_\ga\siml  3\times 10^{-2}$), the models predict X-ray
fluences that ought to be detectable. With  current detectors this is possible
{\it
if} the GRB happens to be in the (usually rather  narrow) X-ray detector
field of view. The latter, unfortunately, is statistically  improbable. For
instance, the Rosat field of view is 3 square degrees, i.e. $\sim 10^{-4}$ of
the entire sky, so
for a burst rate of one per day, one would have to wait $\sim 10^4$ days to
find a GRB in the field of view. This situation, however, should greatly
improve with the advent of a new generation of omnidirectional X-ray detectors,
e.g. HETE (Ricker, \etal, 1993). The predicted X-ray fluences, even when
satisfying the X-ray paucity constraint, can reach values $S_x\simg
10^{-6}-10^{-7}\flu$ for the brighter bursts, which is above  threshold values
for typical detectors, if the source is in the field of view.

The optical/UV fluences, again for sources satisfying the canonical X-ray
paucity ratio, are predicted to be in the range $S_u\simg 10^{-10}\flu$, and in
some cases (e.g. the frozen-in cosmological model) as high as $S_u\simg 3\times
10^{-8}\flu$ for $D_{28}=1$, when $S_\ga\sim 3\times 10^{-6}\flu$. The nearer
(brighter) bursts of $S_\ga\sim 10^{-4}\flu$ therefore could have $S_u\sim
3\times 10^{-6}\flu$, well  above HETE's sensitivity. Other models, e.g. the
piston model satisfying both $\ga$-ray observability and X-ray paucity, would
predict significantly less UV fluence, e.g. for $S_\ga\sim 10^{-4}\flu$ it
predicts $S_u\sim 3\times 10^{-8}\flu$. The HETE O/UV band limiting sensitivity
can also be expressed in U-band magnitudes as $m_{u,m} \sim
-2.5\log_{10}(S_{u,m} \nu_u^{-1} t_L^{-1} /1.9\times 10^{-20}\erg \cmsqi\si\Hz
)\sim 11.$ for a burst of $S_u/t_L\sim 10^{-9}\erg\cmsqi\si$. The U-magnitude
for an arbitrary burst is $m_u= 11. -2.5\log_{10}(S_u/10^{-9} \erg\cmsqi)
+2.5\log_{10} t_L$, where $t_L$ is the observed burst duration in seconds.

The X-ray and UV fluences discussed above are appropriate for the case when a
restriction of the type $S_x/S_\ga\siml 0.03$ is assumed valid, which requires
in general a break at some energy between the $\ga$-ray and X-ray energies.
 From the Band, \etal (1993)  analysis, it appears that many of the bursts
observed with the spectroscopy detector have breaks at energies ranging from
400 to 100 KeV, and some above that. However, a  number of GRB do not show a
break within the detector energy range (down to $\sim 50 \KeV$). This implies
that there could be some bursts which have a higher X-ray fluence than that
mentioned above, e.g., possibly as much as $S_x/S_\ga \sim 0.1-1$.
If the spectral breaks in some bursts occur in the UV, the O/UV fluence
could then be much larger than for those bursts satisfying an X-ray paucity
restriction.

We note that the self-absorption cut-off would, in all our models,
suppress radio-band emission during the burst, especially in ``cosmological''
models. There is, however, the possibility of a faint radio ``afterglow'',
delayed
by a few days in the cosmological case, by which time the
surface area of the fireball has become much larger (cf. Paczynski and Rhodes
1993).
The delay would be shorter for ``galactic'' models, however, and radio flashes
that might be observed could then more plausibly be attributed to coherent
pulsar-type
emission arising from violent magnetospheric effects around a neutron star.

The fluences at all energies depend, of course, on the assumed electron power
law index and the slopes above and below the break $s_a,~s_b$. The
extrapolation of our results for slope values other than those used here is
straightforward, since the break energies and the fluence at these break points
do not change. For the fiducial power per decade slope above the break assumed
here, the GeV fluences are usually comparable to the MeV fluence (when the MeV
band is near or below the main break). The predicted GeV (or higher energy)
fluences could also depend on a high energy cutoff  of the electrons. (Above
photon energies $\simg 1\TeV$ photon-photon opacity effects  are likely to
reduce the fluences to values below those shown in the figures). The EGRET
fluence sensitivity above 1 GeV (Gehrels, \etal, 1991) is of the order
$S_{\GeV} \simg 10^{-5}\flu$ (this is a rough extrapolation from the steady
source  sensitivity for a $5 \times 10^5\s$ integration; for a burst, the
background would need  to be evaluated more carefully). The instrument has a
$\sim 0.6$ sr field of view as  compared to $4\pi$ for BATSE. If the power
slope is indeed 0, some of the brighter (MeV) bursts should be detectable by
EGRET, and indeed they have been (e.g. Schneid, \etal,  1993, Kwok, \etal,
1993).
\bsk
{\it Acknowledgements}: This research has been partially supported through NASA
NAGW-1522, NAG5-2362 and the Royal Society.
\bsk
\ctl{\bf References}
\bsk
\ref Band, D., \etal, 1993, Ap.J., 413, 281
\ref Begelman, M.C., \Mesz, P. and Rees, M.J., 1993, M.N.R.A.S.,
265, L13
\ref Cavallo, G. and Rees, M.J., 1978, M.N.R.A.S., 183, 359
\ref Fishman, G., 1993, in in {\it Compton GRO Observatory}, AIP Conf.Proc.
280,
  eds. M. Friedlander, N. Gehrels and D. Macomb (A.I.P., New York), p. 669
\ref Gehrels, N., Kniffen, D.A. and Ormes, J.F., 1991, in {\it Proc. 22nd Int.
  Conf. on Cosmic Rays}, L. Drury, ed. (University College, Dublin), p...
\ref Goodman, J., 1986, Ap.J.(Lett.), 1986, 308, L47
\ref Hartmann, D., \etal, 1993, to appear in {\it High Energy Astrophysics},
  J. Matthews, ed. (World Scientific).
\ref Hudec, R., \etal, 1992, in {\it Gamma-ray Bursts}, A.I.P. Conf.Proc. 265,
  W. Paciesas and Fishman, G., eds. (A.I.P., New York), p. 323
\ref Hoshino, M., Arons, J., Gallant, Y. and Langdon, B., 1992, Ap.J., 390, 454
\ref Katz, J.I., 1993, Ap.J., in press
\ref Kouveliotou, C., \etal, 1993, Ap.J.(Letters), 413, L101
\ref Kwok, P.W., \etal, 1993, in {\it Compton GRO Observatory}, AIP Conf.Proc.
280,
  eds. M. Friedlander, N. Gehrels and D. Macomb (A.I.P., New York), p.855
\ref Laros, J.G., \etal, 1984, Ap.J. 286, 681
\ref Meegan, L.A., Fishman, G.J., Wilson, R.B., Paciesas, W.S., Brock, M.N.
  Horack, J.M., Pendleton, G.N. and Kouveliotou, C., 1992, Nature, 335, 143.
\ref \Mesz, P. and Rees, M.J., 1992a, Ap.J., 397, 570.
\ref \Mesz, P. and Rees, M.J., 1992b, M.N.R.A.S., 257, 29P.
\ref \Mesz, P. and Rees, M.J., 1993, Ap.J., 405, 278
\ref \Mesz, P., Laguna, P. and Rees, M.J., 1993, Ap.J., 215, 181.
\ref Moskalenko, E.I., \etal, 1992, in {\it Gamma-ray Bursts}, C. Ho, R.
Epstein
  and E. Fenimore, eds. (Cambridge U.P.), p.127
\ref Murakami, T., \etal, 1992, in {\it Gamma-ray Bursts}, C. Ho, R. Epstein
  and E. Fenimore, eds. (Cambridge U.P.), p.239
\ref Narayan, R., Paczynski, B. and Piran, T., 1992, Ap.J.(Letters), 395, L83
\ref Paczy\'nski, B., 1986, Ap.J.(Lett.), 308, L43
\ref Paczy\'nski, B., 1990, Ap.J., 363, 218.
\ref Paczy\'nski, B., 1992, in {\it Gamma-ray Bursts}, eds. Paciesas, W. and
  Fishman, G. (A.I.P. Conf.Proc. 265, New York), p. 144
\ref \Pacz, B. and Rhoades, J., 1993, Ap.J.(Letters), submitted
\ref Piran, T., Shemi, A. and Narayan, R., 1993, M.N.R.A.S., 263, 861.
\ref Rees, M.J., 1967, M.N.R.A.S., 137, 429
\ref Rees, M.J. and \Mesz, P., 1992, M.N.R.A.S., 258, 41P.
\ref Ricker, G., \etal, 1992, in {\it Gamma-ray Bursts}, C. Ho, R. Epstein
  and E. Fenimore, eds. (Cambridge U.P.), p. 288
\ref Ryu, D. and Vishniac, E., 1991, Ap.J., 368, 411.
\ref Schaefer, B., \etal, 1992, Ap.J.(Lett.), 393, L51
\ref Schneid, E.J., \etal, 1993, in {\it Compton GRO Observatory}, AIP
Conf.Proc. 280,
  eds. M. Friedlander, N. Gehrels and D. Macomb (A.I.P., New York), p.850
\ref Shemi, A. and Piran, T., 1990, Ap.J.(Lett.), 365, L55
\ref Usov, V.V., 1992, Nature, 357, 472
\ref Woosley, S., 1993, Ap.J., 405, 273
\bsk
\ctl{\bf Figure Captions}
\bsk
Fig. 1:~~ Cosmological model spectra for a luminosity distance
$D_L=10^{28}\cm$,
$E_o=10^{51}\ergs$, $\theta=10^{-1}$ and $\eta=10^3$. The first three are
the cosmological frozen-in, turbulent and piston models for an external density
$\n=1$ and duration $t_L=5\s$ (top left, top right, bottom left). Identifying
models
by a label (type-distance, $\log E_o,\log\theta,\log\n,\log\xi,\log\ka$), these
are
FC51-1003, TC51-10-63, PC51-10-32.
The full lines are the total spectra, and the dashed lines are the spectral
sub-components discussed in the text.
The fourth spectrum (bottom right, PC51-1n-32) is the cosmological piston for
three
different densities, $\n=~10^{-3}, 1, 10^3$, and durations $t_L=50.,5.,0.5\s$
(left to right).  A similar density dependence also holds for the frozen-in and
turbulent models.
\bsk
Fig. 2:~~ Halo model spectra for a distance $D=50$ Kpc, $\theta=10^{-1}$. The
first three, for $\n=10^{-3}$, are the frozen-in (top left, FH45-1-303)
turbulent (top right, TH41-1-303), and piston models (bottom right,
PH41-1-303), for two durations, $t_L=10,~0.5\s$ (dashed and full lines).
The bottom right figure is the piston halo model PH41-1n03 for densities
$\n=10^{-3},1,10^3$ and durations $t_L=10,1,10^{-1}\s$ (left to right).
\bsk
Fig. 3:~~ Galactic disk model spectra for a distance $D=1$ Kpc,
$\theta=10^{-1}$,
showing for $\n=1$ the frozen-in (top left, FD43-1003), turbulent (top right,
TD39-1003) and piston (bottom left, PD39-1003) models at two durations
$t_L=5,0.2\s$ (dashed, full lines). In the bottom right figure the piston disk
model PD39-1n03 is shown for three external densities $\n=10^3,1,10^{-3}$ and
durations $t_L=2,0.2,0.02\s$ (left to right).
\bsk
Fig. 4:~~ Frozen-in cosmological model FC51-1003, for an external density
$\n=1$
and $D_L=10^{28}\cm$. The model parameters are FC51-1003, giving
(type-distance, $\log E_o,\log\theta,\log\n,\log\xi,\log\ka$).
The lower left flat portion is not part of the spectrum,
it represents the HETE threshold at fluence $F\sim 10^{-10}\flu$. In order of
increasing energies, the breaks are due to \sync and IC in the reverse shock,
and reverse \sync photons IC-scattered in the blast wave.
\bsk
Fig. 5:~~ Turbulent cosmological model TC51-10-63 for $\n=1$, $D_L=10^{28}\cm$.
As before, the detection threshold is at $10^{-10}\flu$, in order of increasing
energy the breaks are due to \sync and IC in the reverse shock, reverse \sync
photons IC-scattered on the blast wave, and blast wave \sync photons
IC-scattered
in the blast (the blast \sync component is submerged by the others).
\bsk
Fig. 6:~~ Turbulent halo model TH41-1-303 for an external density $\n=10^{-3}$
and $D=50$ Kpc. Detection threshold is $10^{-10}\flu$. The main breaks seen are
the reverse \sync, blast \sync and combined IC (barely seen at low durations),
with
increasing energies.
\bsk
Fig. 7:~~ Piston Cosmological model PH51-10-32 for an external density $\n=1$,
and detection hreshold at $10^{-10}\flu$.
The two breaks are due to \sync and IC photons from the blast.
\bsk
Fig. 8:~~ Piston (galactic) disk model PD39-1003, for an external density
$\n=1$, distance $D=$ 1 Kpc and detection threshold $10^{-10}\flu$. The main
break is due to the blast wave \sync photons, while the blast wave IC is at
very
high energies, and relatively weaker.
\bsk
Fig. 9:~~ The $\gamma$-ray band (0.5 MeV) fluence contours and the $\gamma$-ray
to X-ray band fluence ratios (0.5 MeV/1 KeV), in the duration-external density
parameter space, for the three types of model F, T, P (columns) at the three
distance scales C, H, D (rows). The $\gamma$ fluence has three contour levels,
$S_\ga=10^{-8}, 10^{-6}, 10^{-4}\flu$ (dotted, short-dashed and long-dashed
lines,
unlabeled); the $\gamma/X$ ratio has three contour levels, $log S_\ga /\log S_x
=
1., 1.5, 2.$ (dot-dash, dot- long dash, and dotted, with label $r$).
In the halo and galactic disk cases, models below and to the left of the
solid diagonal line have $\eta > \Gamma_{max}=3.3\times 10^2
E_{42}^{1/3}r_6^{-2/3}$
(see text), i.e., they don't satisfy the impulsive approximation; in this
region,
the duration $t_L$ is not given by eq.(2.1.4) as plotted, but must be specified
as an additional free parameter.
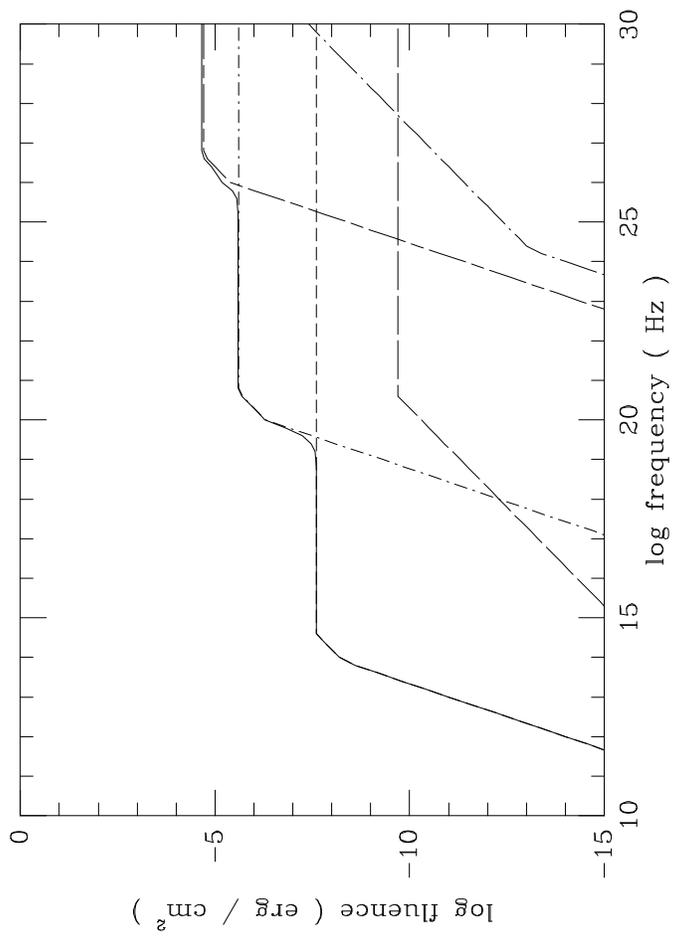
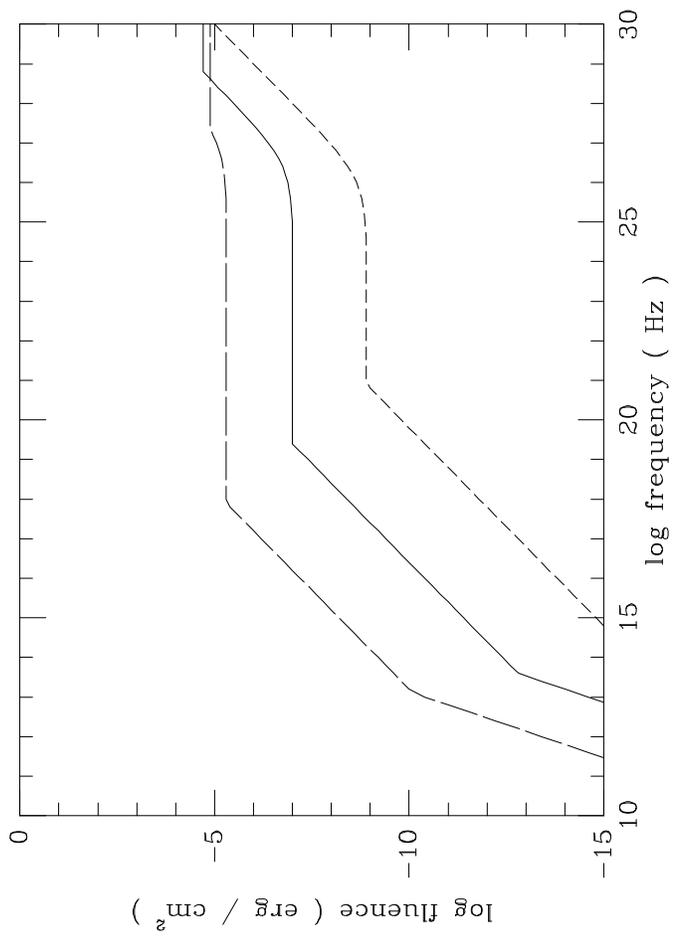
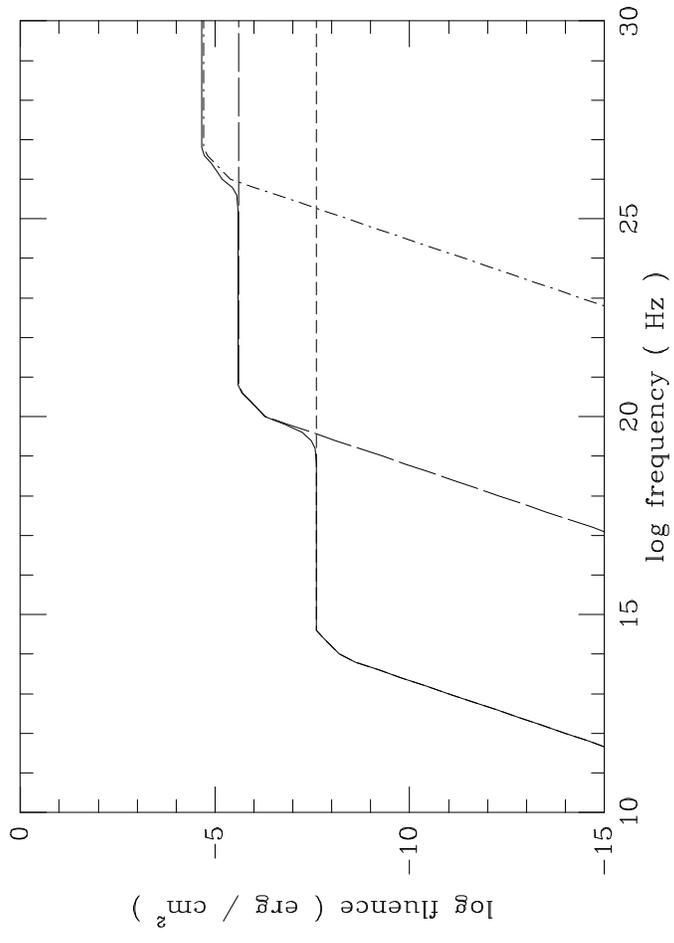
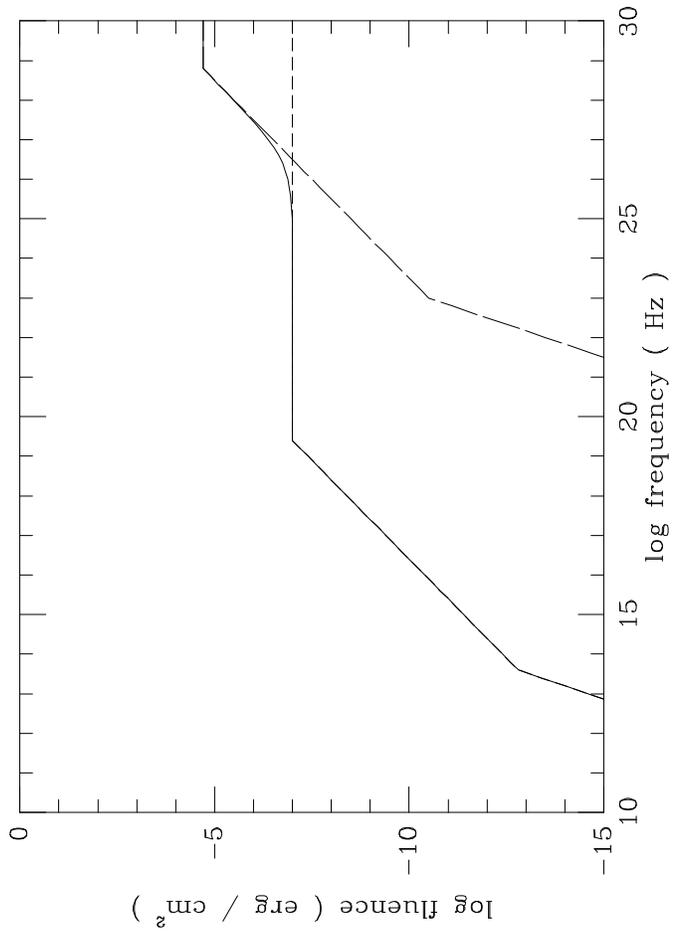
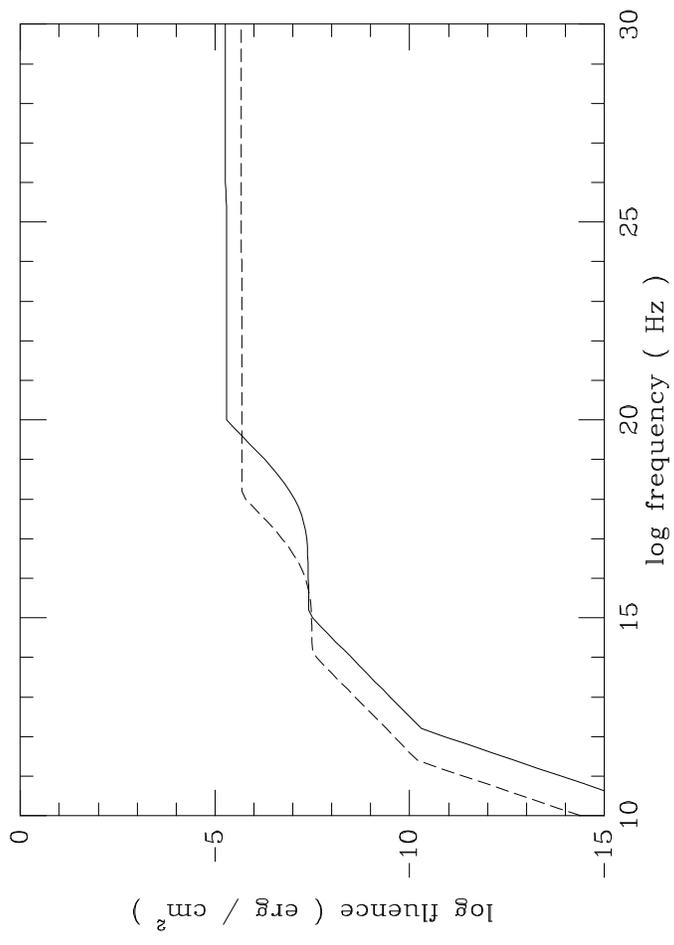
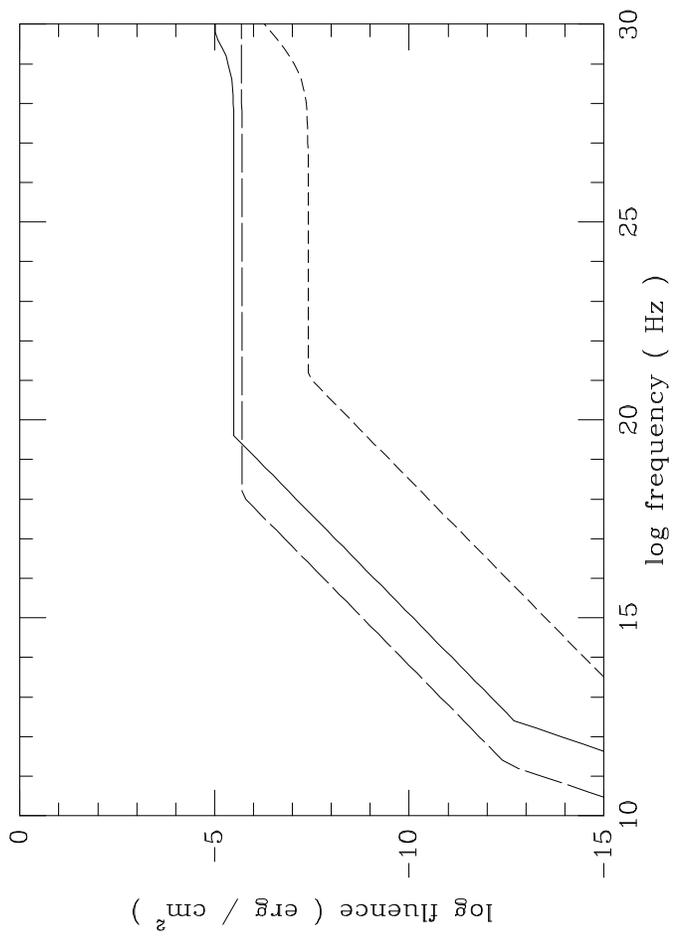
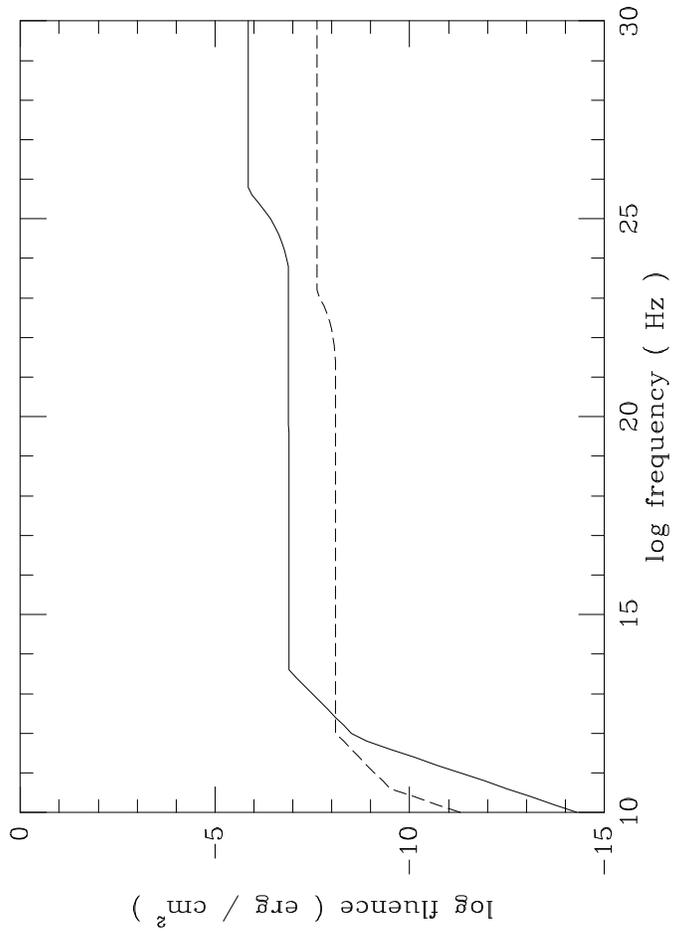
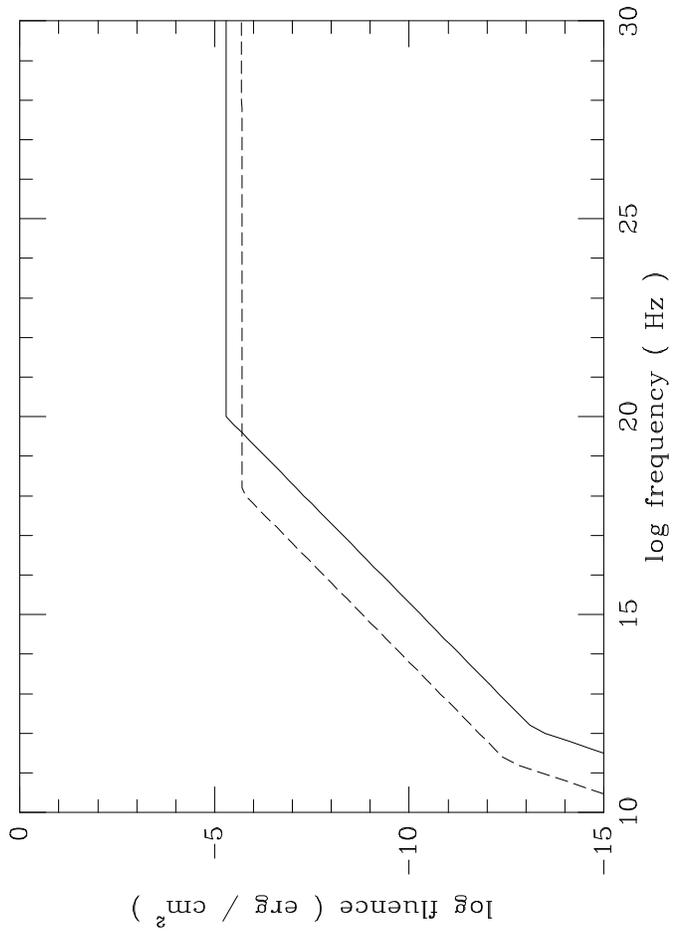
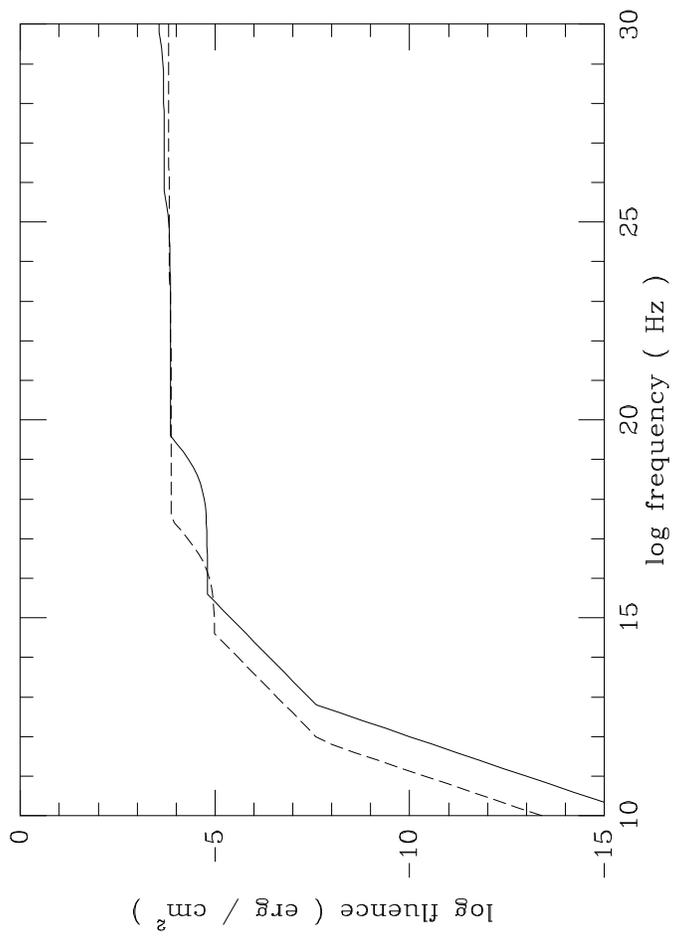
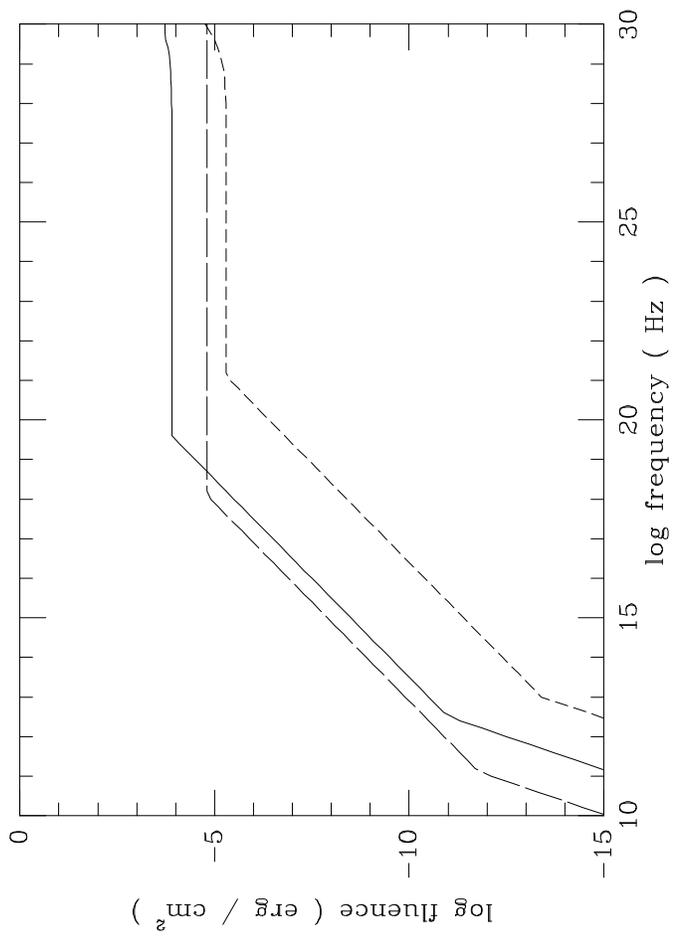
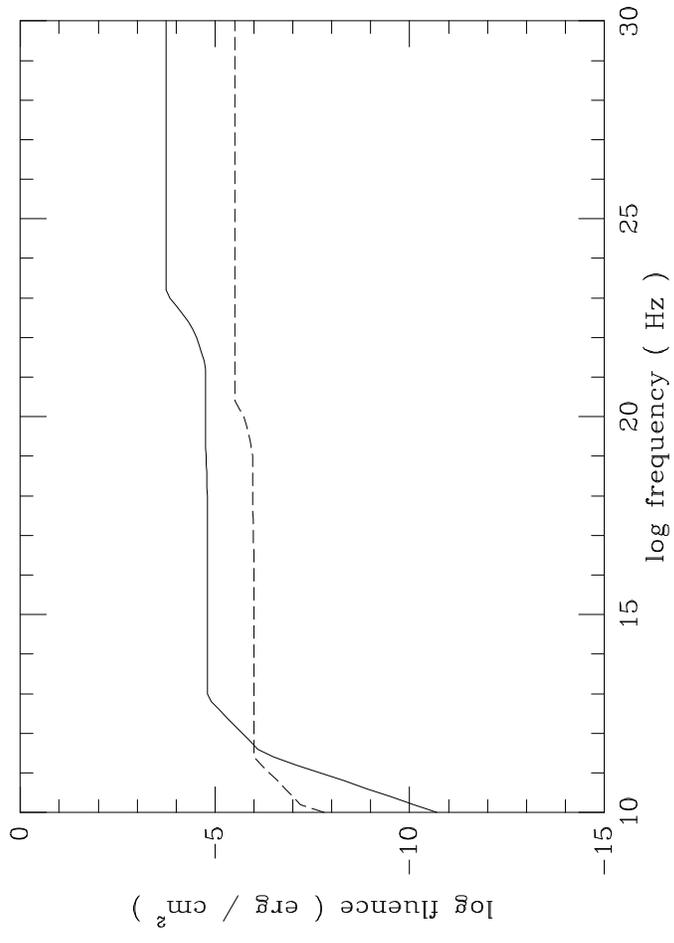
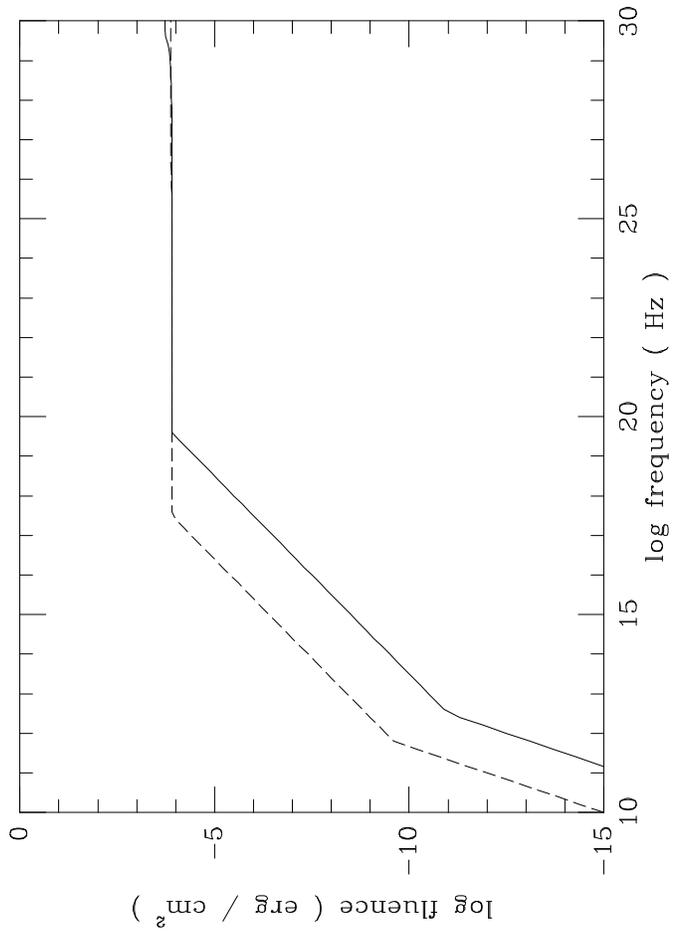
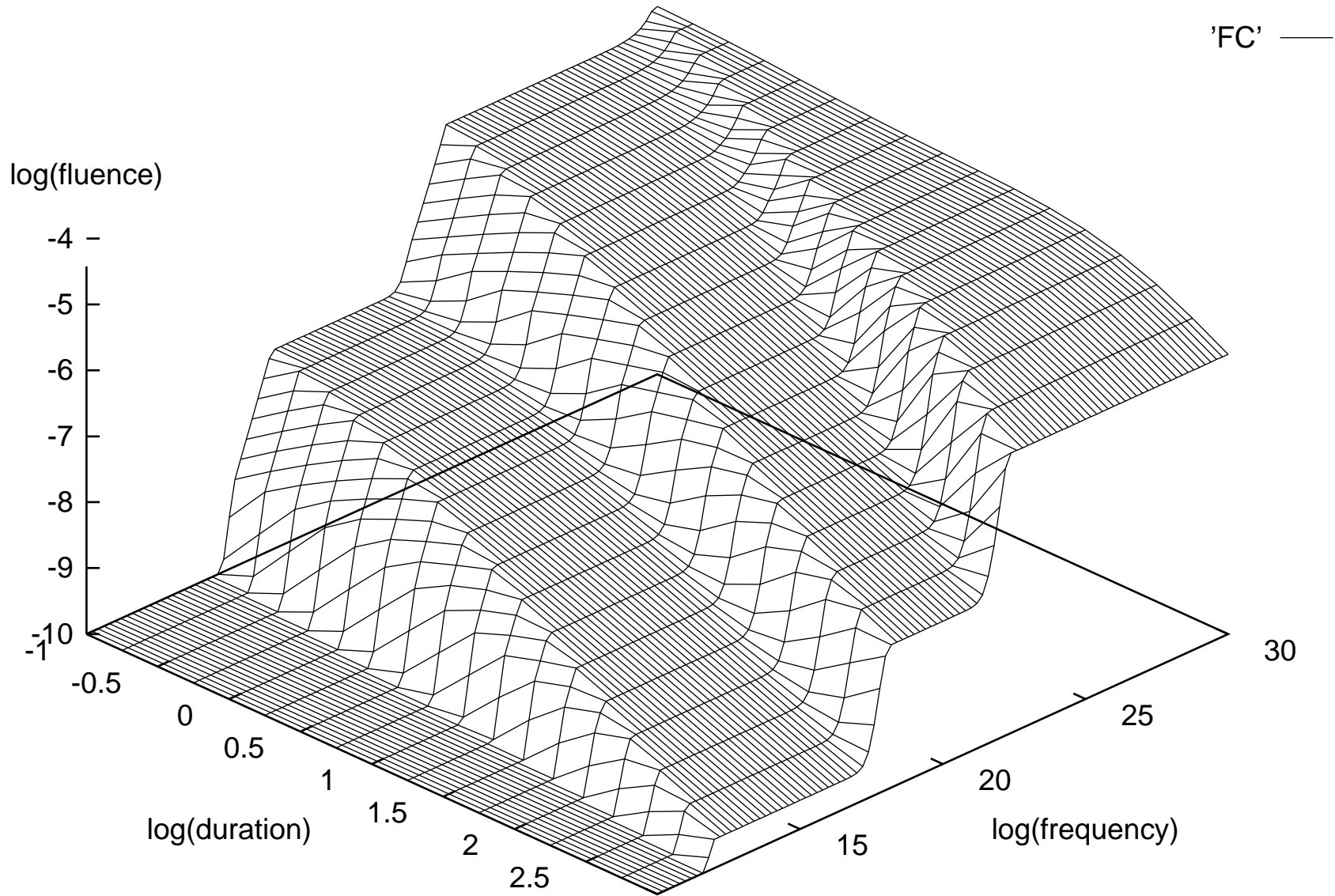
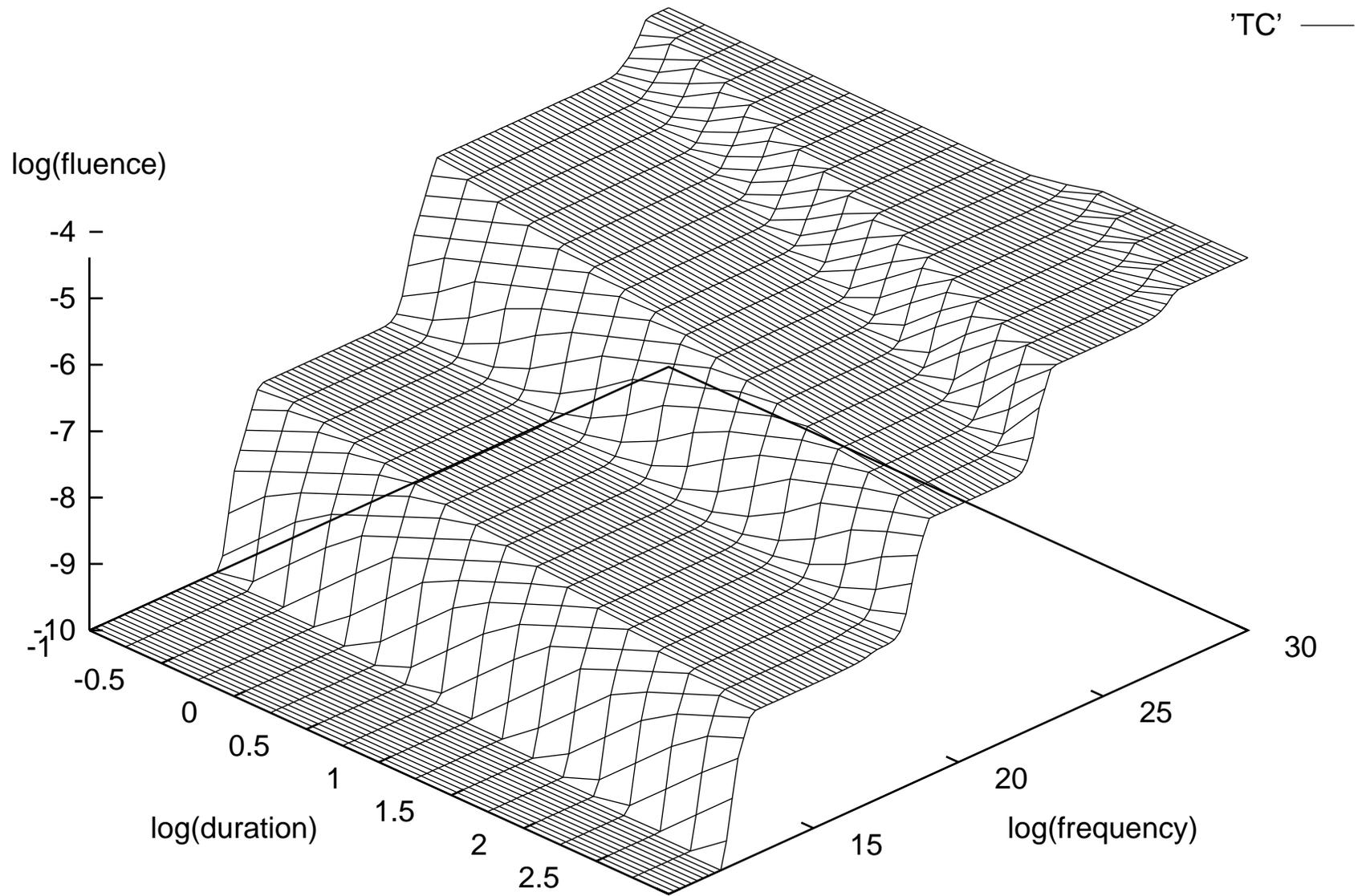
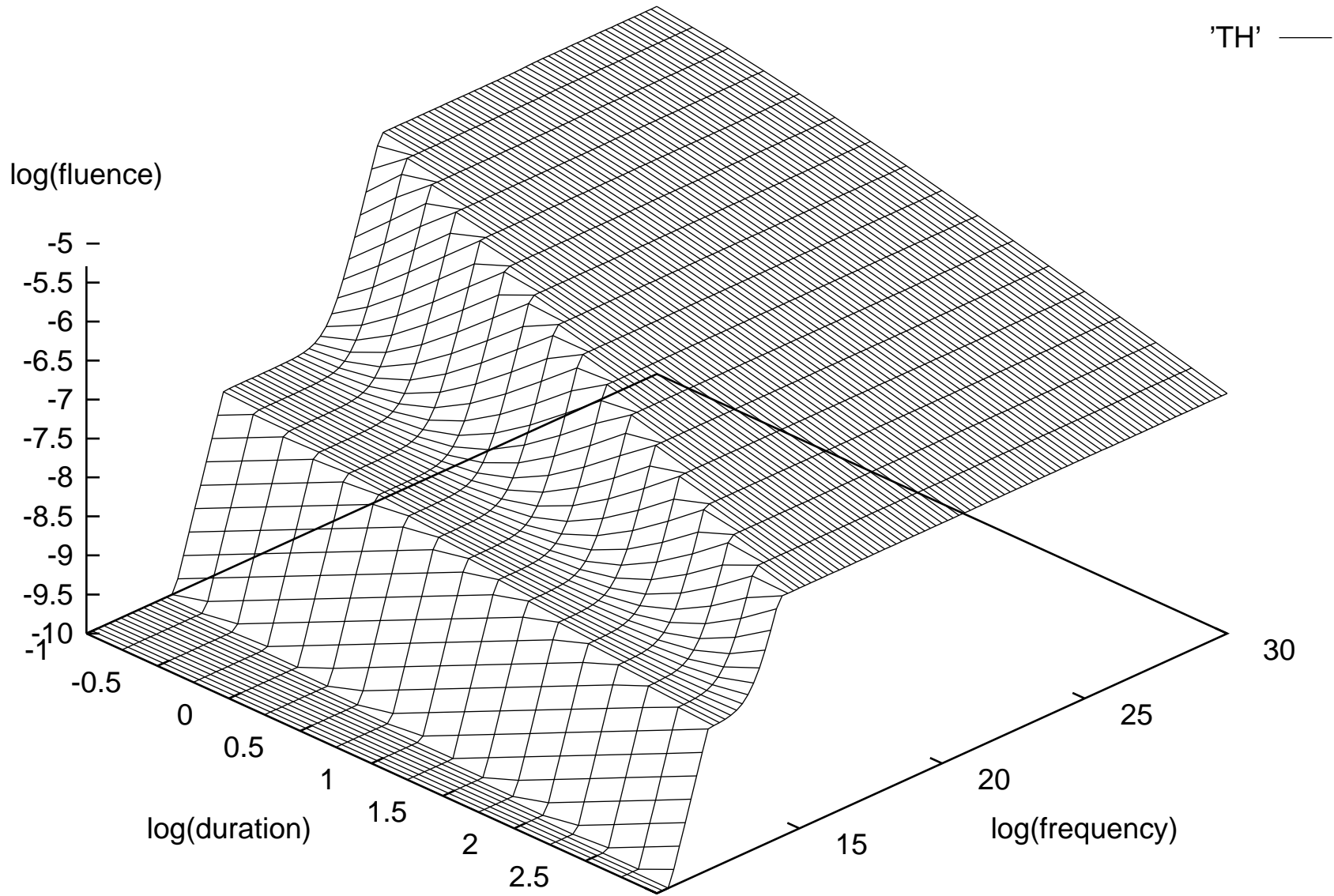
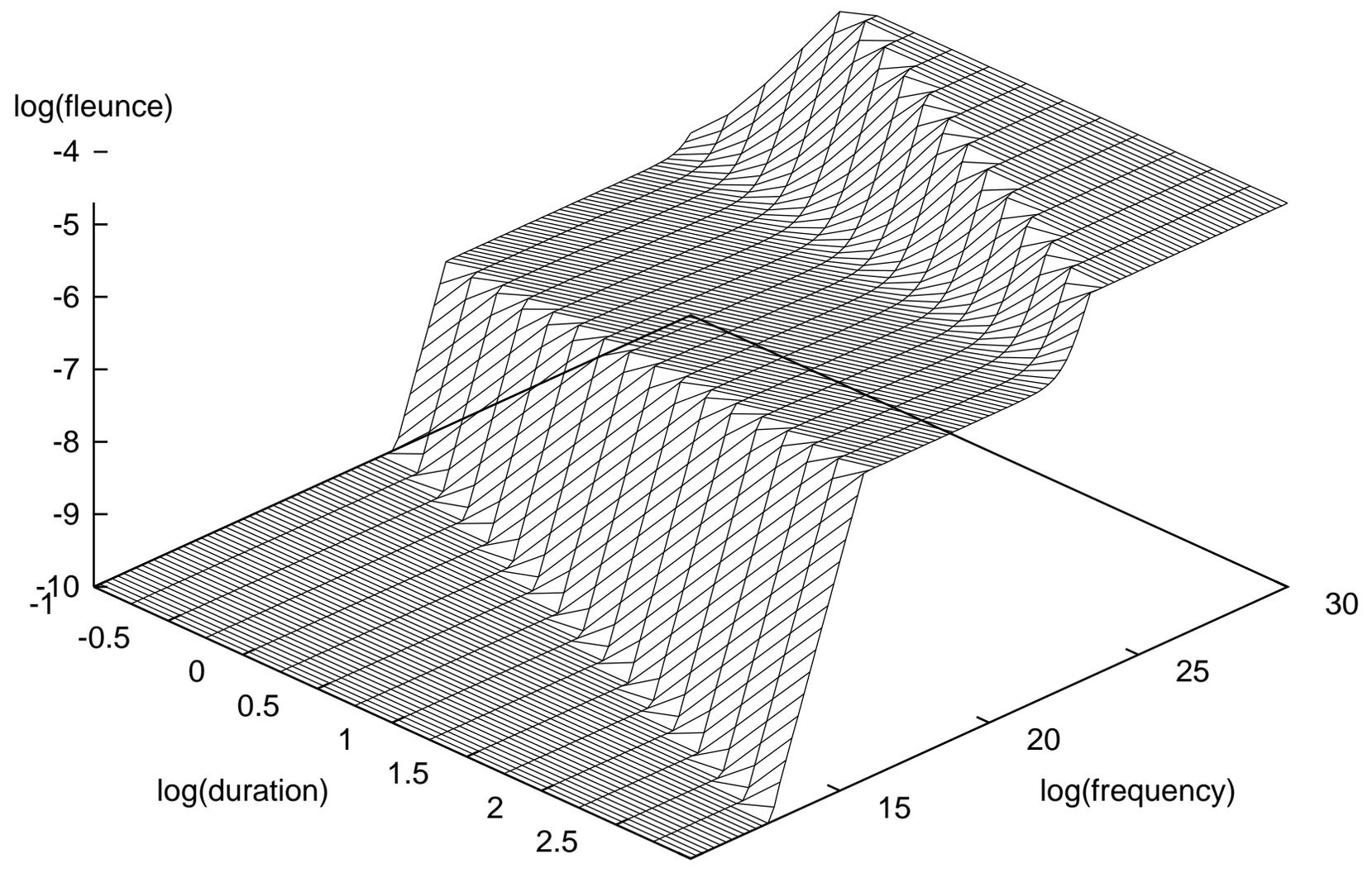
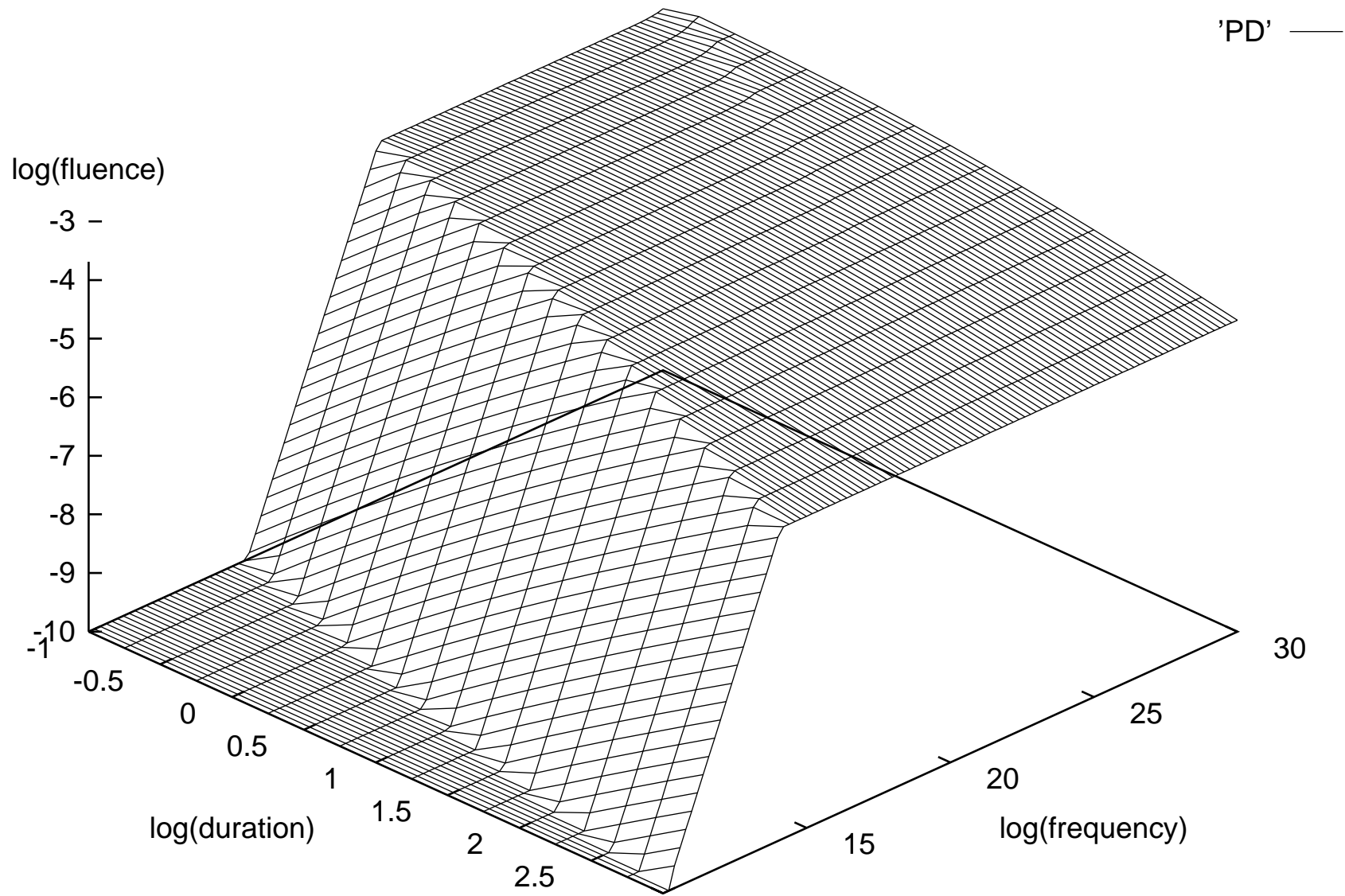
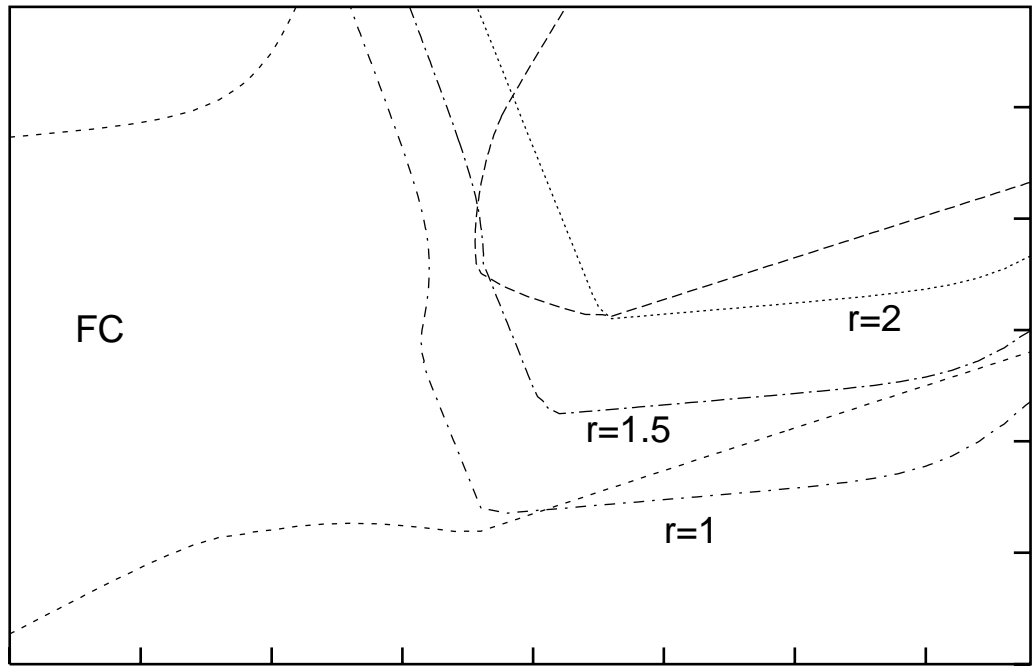
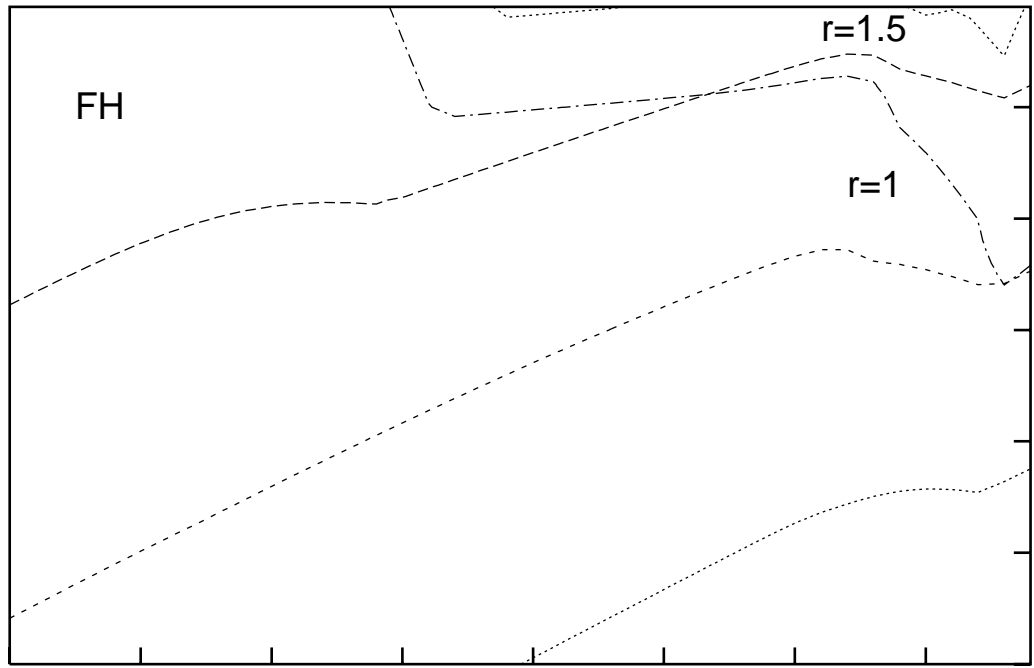
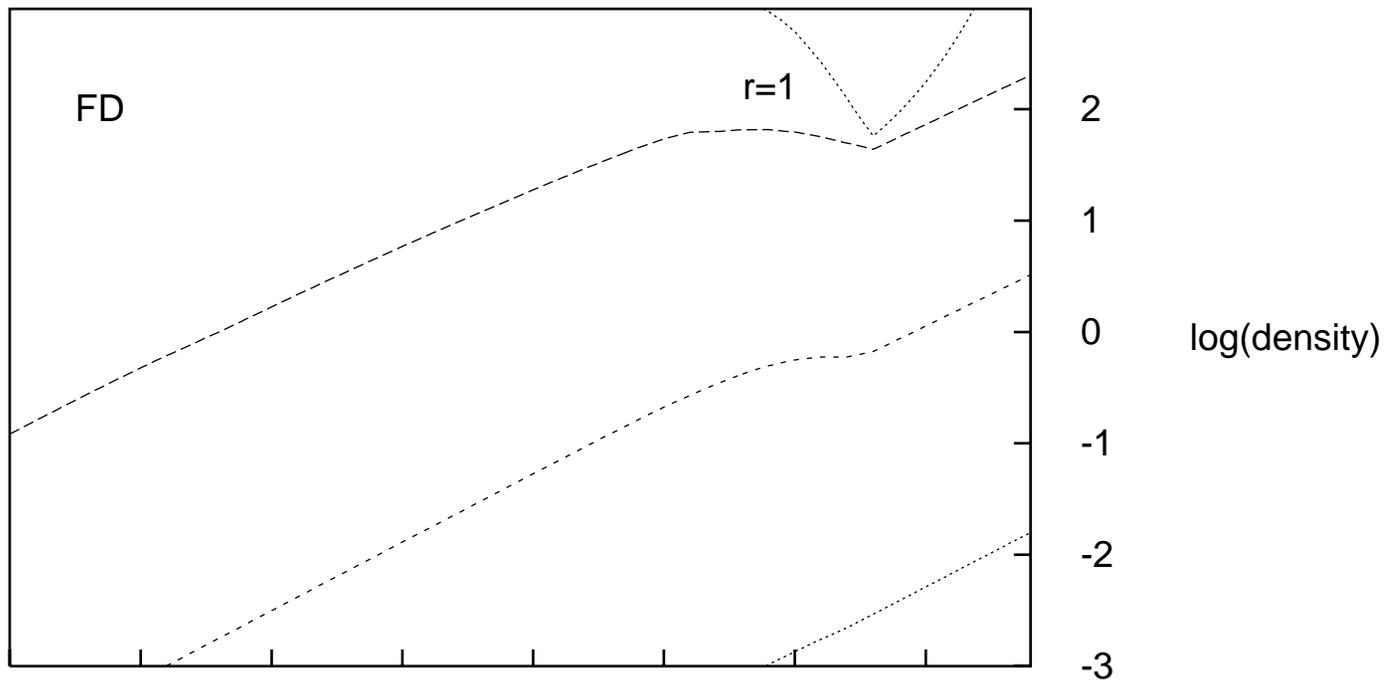
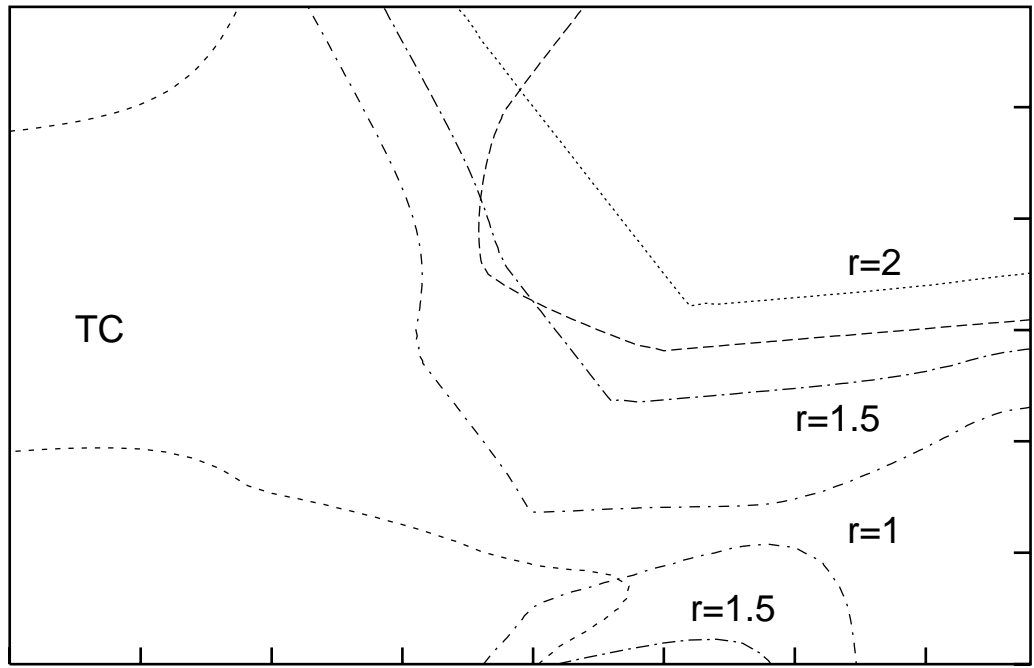
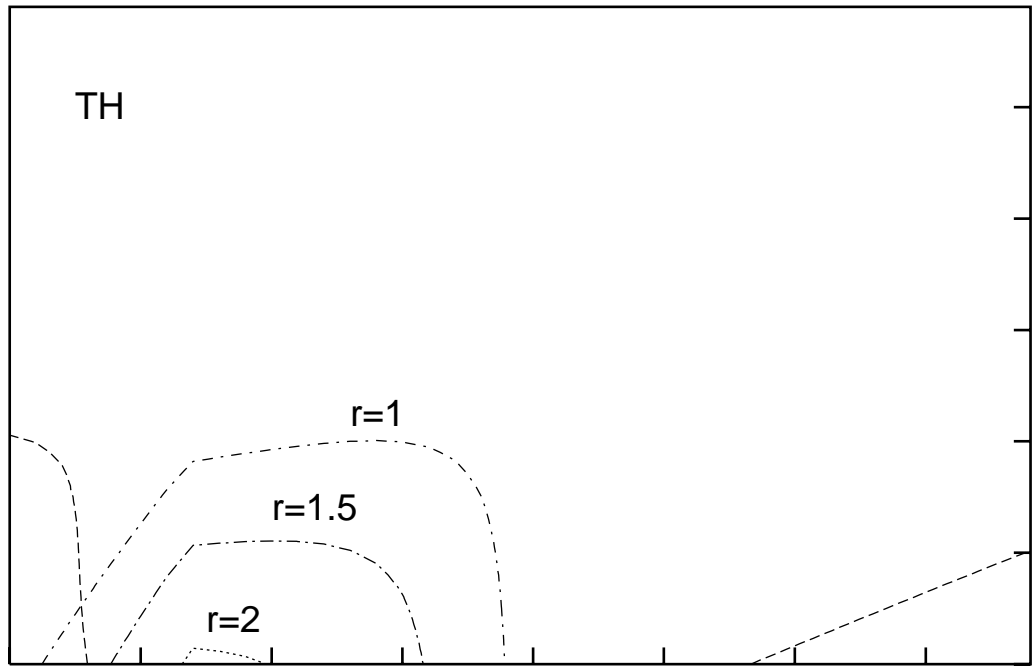
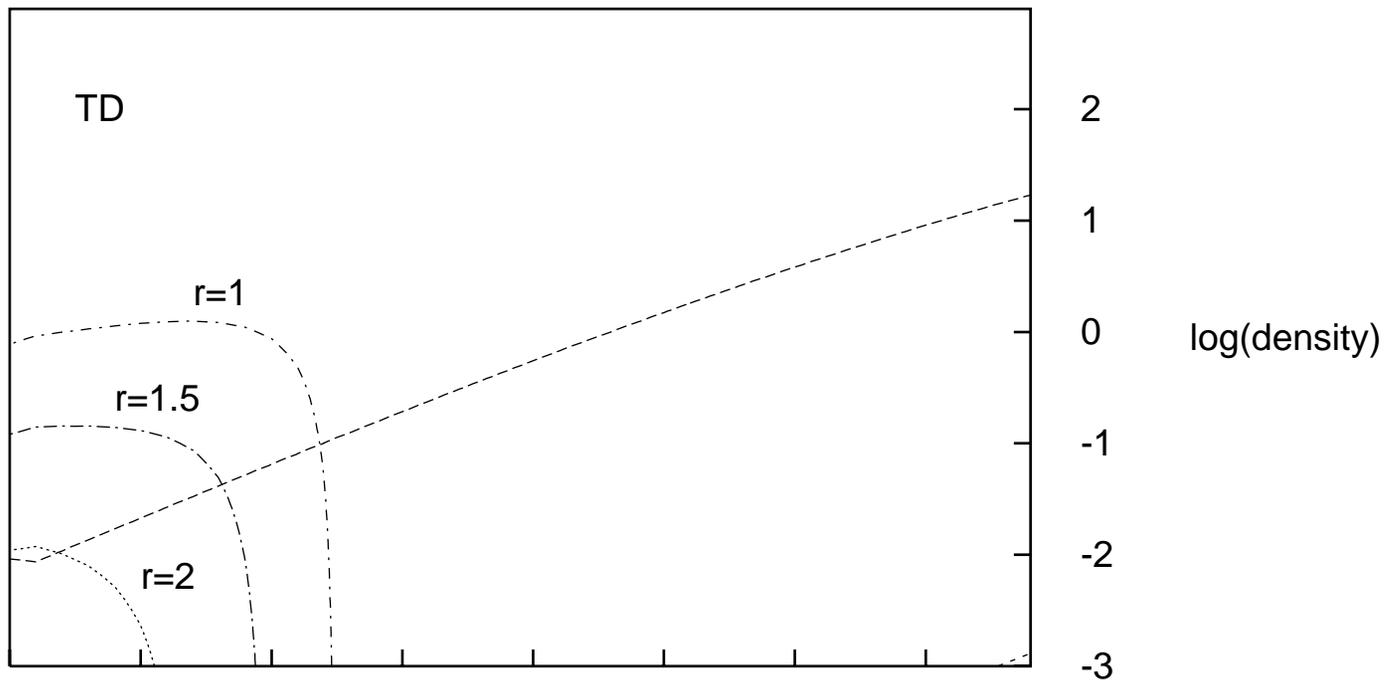
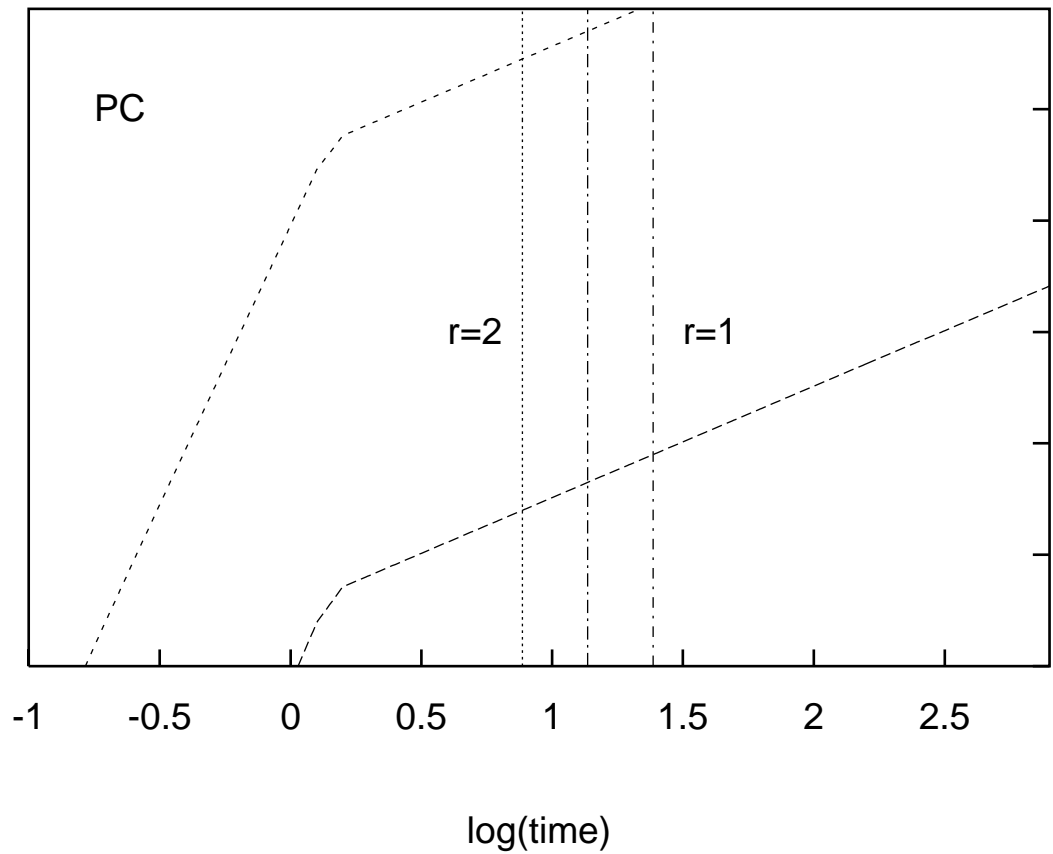
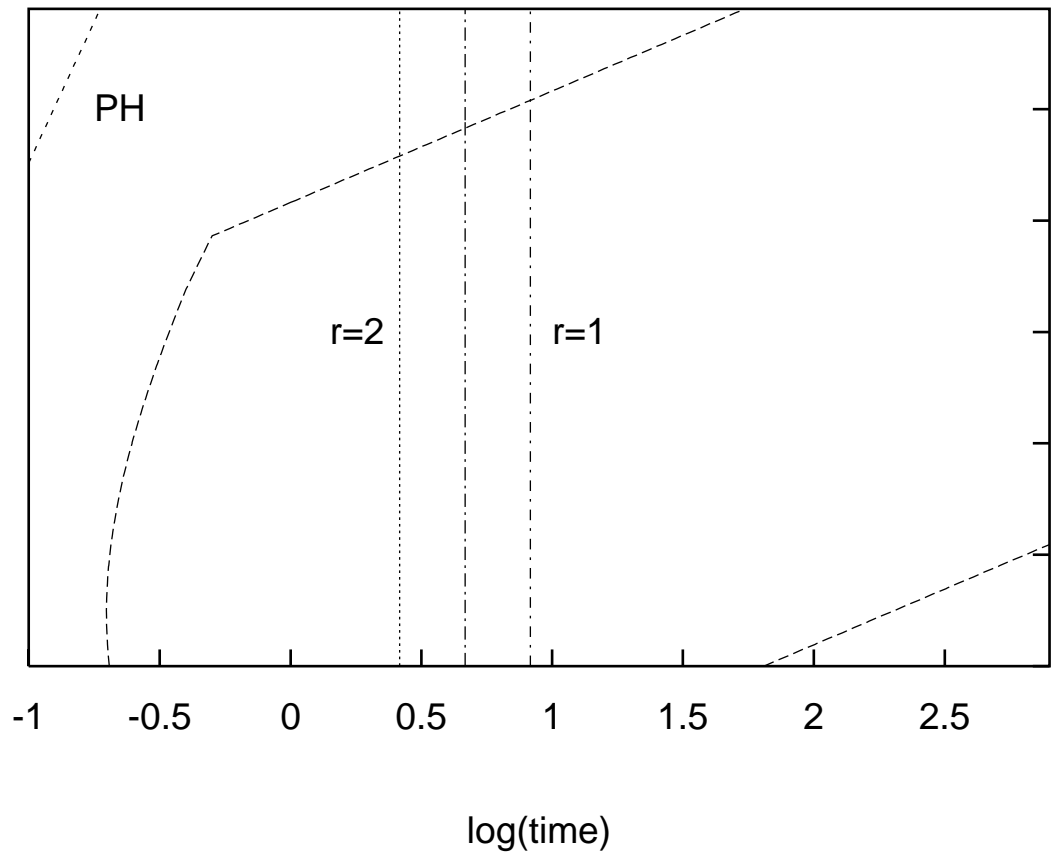
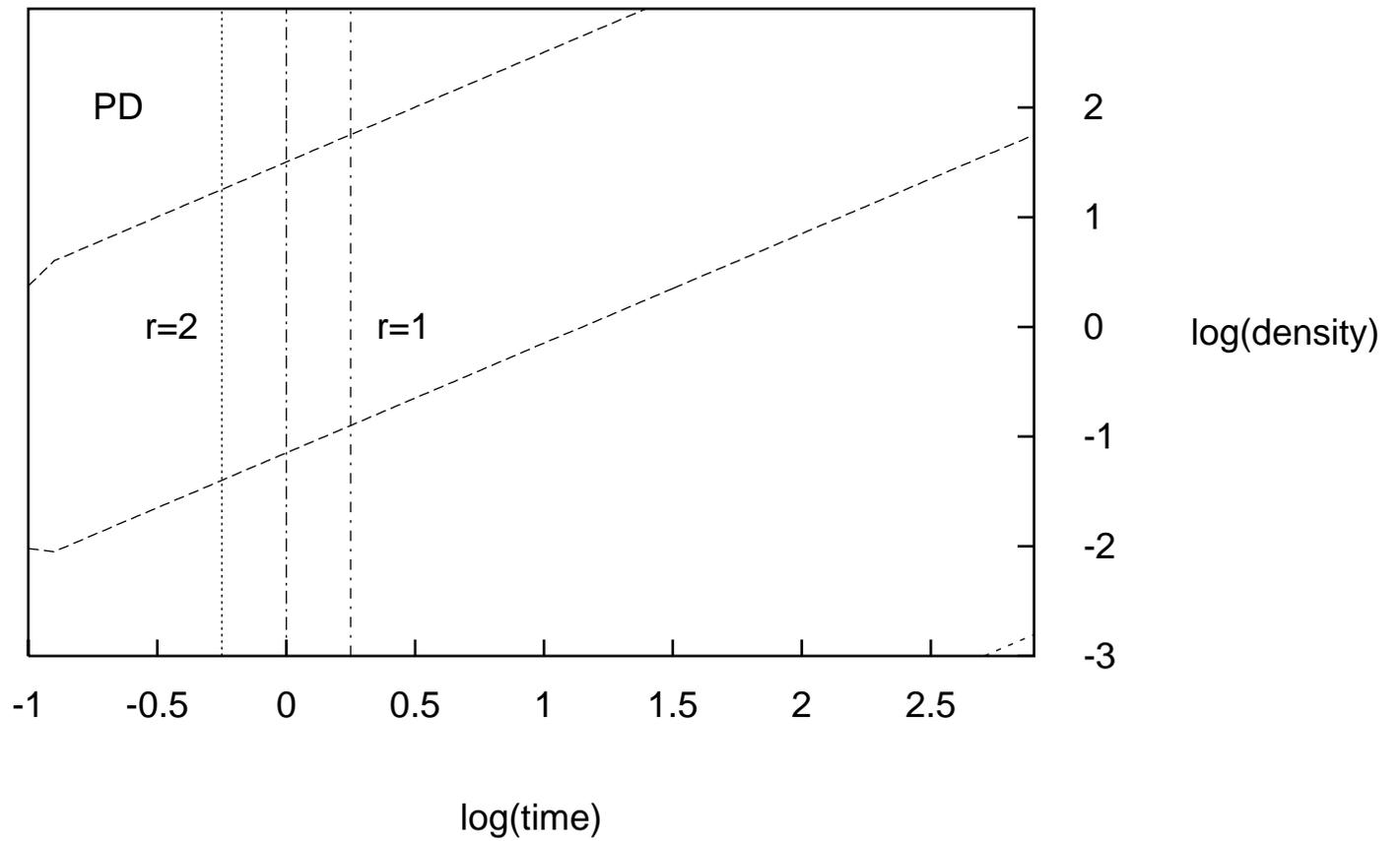
\end